\documentclass[a4paper,fleqn,usenatbib]{mnras}

\usepackage{newtxtext,newtxmath}

\usepackage[T1]{fontenc}
\usepackage{ae,aecompl}

\usepackage{graphicx}	
\usepackage{subfig}
\usepackage{verbatim}

\newcommand{\cii}{{\sc [CII]~}}
\newcommand{\as}{$\,{\rm arcsec}$}
\newcommand{\irxb}{${\rm IRX}$--$\beta$}
\newcommand{\Lsun}{${\rm L}_{\odot}$}
\newcommand{\sfrunit}{$/{\rm M}_{\odot}/{\rm yr}$}

\title[Dust morphology in bright $z = 7$ LBGs]{The discovery of rest-frame UV colour gradients and a diversity of dust morphologies in bright $\mathbf {z \simeq 7}$ Lyman-break galaxies}

\author[R. A. A. Bowler et al.]{R. A. A. Bowler,$^{1}$\thanks{E-mail: rebecca.bowler@physics.ox.ac.uk} F. Cullen,$^{2}$ R. J. McLure,$^{2}$ J. S. Dunlop,$^{2}$ A. Avison$^{3,4,5}$\\
$^{1}$ Astrophysics, The Denys Wilkinson Building, University of Oxford, Keble Road, Oxford, OX1 3RH \\
$^{2}$ SUPA\footnote{Scottish Universities Physics Alliance}, Institute for Astronomy, University of Edinburgh, Royal Observatory, Edinburgh EH9 3HJ, UK\\
$^{3}$ Jodrell Bank Centre for Astrophysics, Department of Physics and Astronomy, School of Natural Sciences, The University of Manchester, Manchester, M13 9PL, UK\\
$^{4}$ UK ALMA Regional Centre Node, Manchester, M13 9PL, UK\\
$^{5}$ SKA Observatory, Jodrell Bank, Lower Withington, Macclesfield, SK11 9FT\\
}

\pubyear{2021}

\begin{document}

\label{firstpage}
\pagerange{\pageref{firstpage}--\pageref{lastpage}}
\maketitle

\begin{abstract}
We present deep ALMA dust continuum observations for a sample of luminous ($M_{\rm UV} < -22$) star-forming galaxies at $z \simeq 7$.  
We detect five of the six sources in the far-infrared (FIR), providing key constraints on the obscured star-formation rate (SFR) and the infrared-excess-$\beta$ (\irxb) relation without the need for stacking.
Despite the galaxies showing blue rest-frame UV slopes ($\beta \simeq -2$) we find that $35$--$75$ percent of the total SFR is obscured.
We find the \irxb~relation derived for these $z \simeq 7$ sources is consistent with that found for local star-burst galaxies. 
Using our relatively high-resolution (FWHM $\simeq 0.7\,{\rm arcsec}$) observations we identify a diversity of dust morphologies in the sample.
We find both compact emission that appears offset relative to the unobscured components and extended dust emission that is co-spatial with the rest-frame UV light. 
In the majority of the sources we detect strong rest-frame UV colour gradients (with up to $\Delta \beta \simeq 0.7$--$1.4$) as probed by the multi-band UltraVISTA ground-based data.
The observed redder colours are spatially correlated with the location of the FIR detection.
Our results show that even in bright Lyman-break galaxies at $z \simeq 7$ the peak of the star-formation is typically hosted by the fainter, redder, regions in the rest-frame UV, which have an obscured fraction of $f_{\rm obs} \ge 0.8$.
As well as demonstrating the importance of dust obscured star-formation within the Epoch of Reionization, these observations provide an exciting taster of the rich spatially resolved datasets that will be obtained from~\emph{JWST} and high-resolution ALMA follow-up at these redshifts. 
\end{abstract}

\begin{keywords}
galaxies: evolution -- galaxies: formation -- galaxies: high-redshift
\end{keywords}



\section{Introduction}
A complete understanding of galaxy evolution requires a measurement of dust-obscured star formation across cosmic time (e.g.~\citealp{Madau2014, Hodge2020}).
Within a star-forming galaxy, the impact of interstellar dust can be observed as a reddening of the rest-frame UV to near-infrared (NIR) emission from stars, as these photons are absorbed and the energy is reradiated in the far-infrared (FIR).
The detailed wavelength dependence and magnitude of this reddening is parameterised as an attenuation curve, which in addition to the extinction of photons due to the dust absorption, also includes scattering into and out of the line-of-sight.
Fundamentally the attenuation curve depends on the physical properties of the dust grains, in particular the composition and size distribution (e.g.~\citealp{Draine2003}) as well as the dust-to-star geometry  (e.g.~\citealp{Calzetti2001}).
On a practical level, an understanding of the degree of attenuation of the star-light within galaxies is essential to determine the total star-formation rate (SFR) of the source.
In particular, when only the rest-frame UV part of the spectral-energy distribution (SED) is observed, a calibration is required to ascertain how much dust obscured star-formation is present.
In local star-burst galaxies, a correlation between the rest-frame UV colour (encapsulated in the rest-frame UV slope, $\beta$; $f_{\lambda} \propto \lambda^\beta$) and the ratio of the FIR to UV luminosity (a proxy for the ratio of the obscured to unobscured SFR) has been found, demonstrating that, in general, redder galaxies have a higher proportion of obscured star formation (e.g.~\citealp{Meurer1999, Takeuchi2012}).
In the absence of data measuring the FIR part of the SED, this local `infrared excess-$\beta$' (IRX-$\beta$) relation has been assumed to hold to higher redshifts and into the Epoch of Reionization (EoR) at $ z > 6.5$.
Given that the colours of galaxies within this epoch appear predominantly blue ($\beta \simeq -2$;~\citealp{Dunlop2013, Rogers2013, Bouwens2014beta}), this has led to the conclusion that dust obscured star-formation was minimal within the first billion years~\citep{Duncan2014, Bouwens2016b}.
Now, with increasingly large samples of high-redshift galaxies being targeted with the unrivalled sensitivity and high spatial resolution of the Atacama Large Millimeter/Sub-millimeter Array (ALMA), it is possible to directly test these assumptions and robustly measure the impact of obscured SF within the EoR for the first time.

Initially, measurements of the FIR dust continuum emission in $z > 6$ galaxies were performed on individual sources or very small samples, typically with the primary purpose of detecting the {\sc [OIII]} $88\,\mu{\rm m}$ or {\sc [CII]} $158\,\mu{\rm m}$ fine structure lines.
These observations have produced a variety of results on the detection of FIR continuum flux, with some studies suggesting that dust obscuration is already established at very high redshifts~\citep{Watson2015, Bowler2018, Tamura2019, Laporte2021}, while other studies reported non-detections~\citep{Ouchi2013, Capak2015}.
In parallel to these targeted observations there have been several efforts to create mosaics using ALMA, building upon deep extragalactic fields where multi-wavelength datasets have already been secured: for example, continuum imaging in Band 3 and 6 in the~\emph{Hubble} Ultra-Deep Field (HUDF) as part of the ALMA deep field~\citep{Dunlop2017} and ALMA Spectroscopic Survey in the HUDF (ASPECS;~\citealp{GonzalezLopez2020}).
By providing a measurement of the flux, or an upper limit, at the position of all galaxies within the field, these mosaics have allowed the dust properties of galaxies to be determined over a range of physical properties and redshifts.
These studies have demonstrated that stellar mass ($M_{\star}$) is the strongest predictor of whether a source will be detected in the FIR, as the most massive sources also show the highest total SFR and obscured SFR fraction~\citep{Dunlop2017, McLure2018, Bouwens2020}.

Due to the small field of view of ALMA however, these mosaics have only captured a handful of the rare, massive and star-forming end of the galaxy population at very high redshift ($z > 6$), as such sources have surface densities of $\sim 10 \,/{\rm deg}^2$.
In response, there have been a succession of ALMA large programs undertaken to provide targeted observations covering the first statistical samples of massive $z \gtrsim 6$ sources.
The ALMA Large Program to Investigate C+ at Early Times (ALPINE;~\citealp{LeFevre2020, Bethermin2020, Faisst2020}) survey focused on detecting {\sc [CII]} and also the dust continuum in a sample of $\sim 120$ spectroscopically confirmed galaxies at $4.4 < z < 5.8$ with SFR $= 10$--$100\,{\rm M}_{\odot}/{\rm yr}$ and $M_{\star} > 10^{9}\,{\rm M}_{\odot}$.
At even higher redshifts, this work has been extended by the Reionization Era Bright Emission Line Survey (REBELS;~\citealp{Bouwens2021}) to include 40 galaxies with photometric redshifts in the range $z = 6.4$--$9.5$.
Even within these impressive datasets, detections of the dust continuum have proved elusive, with ALPINE only detecting 23 of the 120 targeted galaxies at $> 3.5\sigma$ significance~\citep{Fudamoto2020} and a $< 40\,{\rm percent}$ detection rate in REBELS~\citep{Bouwens2021}.
This lack of direct constraints on the dust continuum emission means that the calibration of the IRX-$\beta$ relation has to be based predominantly on stacking, which given the large errors on the $\beta$-slope measurement, can lead to biases (e.g.~\citealp{McLure2018}).
Indeed, several recent stacking analyses have come to opposite conclusions about the degree of dust-obscured star formation, with some finding consistency with the local Meurer relation~\citep{Fudamoto2019, McLure2018} and others finding a deficit of FIR flux for a given UV colour~\citep{Bouwens2016, Fudamoto2020, Schouws2021}.

In the expectation of a relationship between the observed IRX and the rest-frame UV slope there is the assumption that the stars and dust are well mixed, which leads to the coupling of any observed reddening in the UV to the FIR emission detected (e.g.~\citealp{Meurer1999, Charlot2000, Calzetti2001}).
If instead the galaxy consists of regions of significantly different obscuration, then the relationship will break down for the galaxy as a whole.
Indeed, geometric effects have been put forward as an explanation of the discrepant results at $z > 5$~\citep{Faisst2017, Popping2017}.
In local starburst galaxies, the existence of an \irxb~relation and a clear morphological similarity, indicates that the rest-frame UV emission from young stars is being attenuated from dust that is tracing broadly the same star-forming regions of the galaxy (e.g. in the spiral arms;~\citealp{Kennicutt2003, GildePaz2007}).
In the high-redshift Universe however, where galaxies become more turbulent and irregular (e.g.~\citealp{ForsterSchreiber2011, Buitrago2013, Guo2015}) the expected morphology of the dust relative to the observed UV emission is not clear.
Evidence for offset dust continuum emission relative to the rest-UV has been identified in several high-redshift LBGs~\citep{Koprowski2016, Laporte2017, Faisst2017, Bowler2018}, and similar trends have been found when comparing the {\sc [CII]} FIR line and the rest-UV continuum (e.g.~\citealp{Maiolino2015, Carniani2017}).
Although some of these offsets have been attributed to astrometric systematics (e.g.~\citealp{Dunlop2017}) there is a growing consensus that FIR continuum and line emission are frequently physically offset as compared to the observed rest-UV emission (see~\citealp{Carniani2018}).
Whether high-redshift galaxies show large and distinct regions of obscured and unobscured star-formation has implications for the use of the IRX-$\beta$ relation in deriving the cosmic SFR density (e.g.~\citealp{Bouwens2016b}), as the assumed energy balance will break down~\citep{Buat2019} and the global $\beta$ measurement will not be representative of the full source.

A key challenge in identifying offsets between the rest-frame UV and FIR emission at high redshift is that the majority of studies to-date have been primarily detection experiments, which has led to the spatial resolution of the data being fairly low (full-width at half-maximum, FWHM $\gtrsim 1\,{\rm arcsec}$).
Furthermore, `typical' $L \simeq L^*$ galaxies (with luminosities around the characteristic luminosity or knee in the luminosity function, $L^*$) at $z > 6$ are known to be extremely compact in the rest-frame UV.
Studies of the size-magnitude relation have shown that even the brightest sources found within the~\emph{Hubble Space Telescope} (\emph{HST}) Cosmic Assembly Near-infrared Deep Extragalactic Survey (CANDELS;~\citealp{Grogin2011, Koekemoer2011}) are barely resolved with the Wide Field Camera 3 (WFC3), with half-light radii of $< 0.2\,{\rm arcsec}$ or $1\,{\rm kpc}$~\citep{Grazian2012, Curtis-Lake2016}.
These small sizes preclude any detailed analysis of the galaxy geometry with respect to the dust, colour gradients or possible `\emph{HST}-dark' components (e.g.~\citealp{Fudamoto2021}).
Gravitationally lensed sources can offer one solution, however uncertain source-plane reconstruction and the low detection rates of dust from these lower-metallicity sources have made this challenging (e.g.~\citealp{Laporte2021}).
An alternative method is to observe extremely luminous and hence highly star-forming sources, which are by their nature more efficient to follow-up than `typical' galaxies at these redshifts (e.g.~\citealp{Capak2015, Bowler2018}).
Crucially, $L >> L^*$ galaxies have been shown to be highly extended in the rest-frame UV, showing multiple clumps of star-formation~\citep{Bowler2017}, as well as being detected more frequently in the FIR with ALMA~\citep{Bowler2018, Schouws2021}.

In this study we present new deep (120 minutes; $\sim 10\mu{\rm Jy}$ RMS) and relatively high-resolution (FWHM $\simeq 0.7\,{\rm arcsec}$) follow-up ALMA observations of the dust continuum emission in five UV-bright $z = 7$ galaxies.
Including archival observations for an additional source leads to a final sample size of six $z \simeq 7$ LBGs with $M_{\rm UV} < -22.4$.
These observations allow us to detect the majority of the sources in the FIR, providing a highly complete ($5/6$) view of the dust continuum with which to fully understand the obscured fraction of star-formation within this population.
In addition, the spatial resolution of our data allows us to identify a dust morphologies and offsets with respect to the rest-frame UV emission.
The structure of this paper is as follows.
In Section~\ref{sect:data} we describe our sample selection and present the new ALMA observations in addition to the archival data that we use.
We present the methods and results in Sections~\ref{sect:methods} and~\ref{sect:results}, respectively.
We end with a discussion of the impact of our results in Section~\ref{sect:diss} and we present our conclusions in Section~\ref{sect:conc}.
Throughout this work we quote magnitudes in the AB system~\citep{Oke1974,Oke1983}.
The standard concordance cosmology is assumed, with $H_{0} = 70 \, {\rm km}\,{\rm s}^{-1}\,{\rm Mpc}^{-1}$, $\Omega_{\rm m} = 0.3$ and $\Omega_{\Lambda} = 0.7$.
At $z = [6.5, 7.2] $ this results in 1 arcsec corresponding to a physical distance of $[5.46, 5.14]\,{\rm kpc}$.

\section{Data and sample}\label{sect:data}

The six galaxies that are considered in this study were initially selected as part of a search for bright $z \simeq 7$ galaxies from ground-based survey data.
Specifically, we utilized an area of $\sim 2\,{\rm deg}^2$ covered by deep optical and near-infrared imaging in the well-studied Cosmological Evolution Survey (COSMOS;~\citealp{Scoville2007}) and UK Infrared Telescope Infrared Deep Sky Survey (UKIDSS) Ultra Deep Survey (UDS;~\citealp{Lawrence2007}) fields.
The result of this search, presented in~\citet{Bowler2012, Bowler2014}, was a sample of $\sim 30$ galaxies with rest-frame UV absolute magnitudes of $-23.1 < M_{\rm UV} < -21.5$ and photometric redshifts in the range $6.5 < z < 7.2$.
Of this sample, the six brightest sources ($M_{\rm UV} \le -22.4$) were observed using ALMA in Cycle 3 (PID: 2015.1.00540.S) with 10 minutes of follow-up per source in Band 6.
The result of these observations, presented in~\citet{Bowler2018}, was the detection of source ID65666 at $5\sigma$ significance and the derivation of upper limits on the other five sources of $< 92$--$139 \, (2\sigma)\, \mu {\rm Jy}$.
A tentative detection from a stacking analysis indicated that the average flux of the undetected sources was $100 \pm 50\,\mu {\rm Jy}$.

In this work we build upon these previous results using significantly deeper observations of all six sources.
The source coordinates and redshifts are shown in Table~\ref{table:ra}.
For the five previously undetected objects, we obtained additional follow-up observations with ALMA in Cycle 6 from the C graded proposal PID: 2018.1.00933.S (PI Bowler).
Each target was observed for a total of 2 hours on source in Band 6, centred on $233\,{\rm GHz}$, to provide a measurement of the rest-frame FIR continuum emission ($\lambda_{\rm rest} \sim 170 \mu {\rm m}$).
We requested a depth of RMS $=10\,\mu{\rm Jy}$ (previous Cycle 3 data requested RMS = $50\,\mu{\rm Jy}$), however the program was only partially observed.
The data were taken in 12 executions between 20-Mar-2019 and 09-Apr-2019.
A total of between 46 and 50 antennas were used in each observation block, with a minimum baseline of 15m and a maximum of between 360m and 500m. 
The calibrated data provided by ALMA were used for this analysis. 
These data were reduced using the CASA ALMA calibration pipeline (Pipeline version 'Pipeline-CASA54-P1-B'). 
Imaging was conducted in CASA version 5.6~\citep{McMullin2007} using the CASA task \texttt{tclean}. 
The pixel RMS for each dataset was determined in the central region within an annulus of radius of $[1.0, 4.0]$ arcsec.
For the naturally weighted data from our new Cycle 6 program we derived an RMS of $13 \pm 1 \,\mu{\rm Jy}$ for each galaxy/pointing, which is to be expected given that we received approximately $2/3$ of the requested integration time.
The new Cycle 6 data has the requested higher resolution leading to a beam size of $0.79 \times 0.62$ arcsec.
This is to be compared to the original Cycle 3 data, which had a beam of $1.36 \times 1.04$ arcsec.

\begin{table}\caption{The sample of six $z = 7$ star-forming galaxies studied in this work.
The first column presents the ID number as defined from previous studies of these galaxies~\citep{Bowler2014, Bowler2017, Bowler2018}.
We then present the R.A. and Dec. derived from the {\sc SExtractor} centroid in the $J+H$ UltraVISTA DR4 imaging.
Five of the sources have spectroscopic redshifts obtained by~\citet{Hashimoto2019},~\citet{Schouws2021},~\citet{Bouwens2021} and Schouws et al. in preparation.
In Column 4 we present these redshifts in addition to the best-fit photometric redshift for object ID279127.
The alternative ID numbers for each source, as defined by the spectroscopic follow-up studies, are shown in the final column.
The sources are ordered in decreasing redshift, and we present the sources in this order in later figures and tables.
}\label{table:ra}
\begin{tabular}{ccccc}
\hline
ID & R.A. & Dec. & $z$ & Alt. ID \\
\hline
65666 & 10:01:40.69 & +01:54:52.49 & 7.152 &  B14-65666 \\
304416 & 10:00:43.37 & +02:37:51.66 & 7.0611 & Z-001 \\
238225 & 10:01:52.30 & +02:25:42.32 & 6.984 & REBELS-30 \\
304384 & 10:01:36.85 & +02:37:49.22 & 6.686 & REBELS-29 \\
169850 & 10:02:06.47 & +02:13:24.09 & 6.633 & REBELS-34 \\
279127 & 10:01:58.48 & +02:33:08.48 & $6.56^{+0.07}_{-0.06}$ & REBELS-31 \\
\hline
\end{tabular}
\end{table}

In addition to these new data we also include archival observations for object ID65666.
We include the original Cycle 3 data where this source was originally detected in the dust continuum, and the deeper and higher resolution Band 6 data obtained from program 2016.1.00954.S in Cycle 4 (PI Inoue).
In Table~\ref{table:depths} we present a summary of the different ALMA observing programs that we utilize in this work.
Note that four of the sources are also included in the REBELS large program (see Table~\ref{table:ra}), however with a variable integration time as part of the search for {\sc [CII]}~\citep{Bouwens2021}.
Furthermore, object ID304416 was included in the study by~\citet{Schouws2021}, with 78 minutes of integration in Band 6 with a derived beam of $1.47 \times 1.21$ arcsec.
We do not include these data in this work due to the relatively poor spatial resolution.

We performed $uv$-tapering of the datasets to investigate flux on wider spatial scales and to allow a closer comparison between different observing programs.
The beam size and depths calculated for the different datasets are presented in Table~\ref{table:depths}.
The different tapers can be split into a `high-resolution' dataset used primarily to investigate the morphology and offsets where spatial resolution is key, and a  `low-resolution' dataset that was used to define whether or not a source was detected.
To create this lower resolution imaging we tapered the new Cycle 6 data using a scale of 0.5 arcsec to provide images of the same resolution as the Cycle 3 imaging from~\citet{Bowler2018}.
For the higher resolution dataset we tapered the Cycle 4 data for ID65666 (naturally weighted beam of $0.29 \times 0.23$ arcsec) to a resolution comparable to the five other sources.
For this data we excluded the channels effected by the \cii emission~\citep{Hashimoto2018}.
The $uv$-tapering was performed using the \texttt{tclean} task in CASA, by iteratively identifying and masking flux from the central and serendipitous sources~\citep{Bowler2018}.
For the Cycle 4 data for object ID65666 we used CASA version 4.7, while for our data we used version 5.6.

\subsection{Photometric and spectroscopic redshifts}

A total of five of the six brightest $z \simeq 7$ sources identified in~\citet{Bowler2012, Bowler2014} are now spectroscopically confirmed via the detection of one or more FIR lines with ALMA.
The spectroscopic redshifts are shown in Table~\ref{table:ra}.
The first of the six sources to be spectroscopically confirmed was ID65666.
A detection of {\sc [CII]} and {\sc [OIII]} was presented in~\citet{Hashimoto2018}, leading to a spectroscopic redshift of $z = 7.152$.
Recently, ID304416 has also been confirmed via \cii emission at a redshift of $z = 7.06$ by~\citet{Schouws2021}.
The remaining four sources were targeted as part of the REBELS survey~\citep{Bouwens2021}, which included spectral line scans for \cii and provides robust spectroscopic redshift measurements for an additional three objects (Schouws et al. in preparation).
\citet{Fudamoto2021} presented a detailed analysis of ID304384/REBELS-29, where they also identified a `UV-dark' companion galaxy at the same redshift, as derived from \cii emission.
For all sources the spectroscopic redshift was consistent with the previously derived photometric redshift utilised in~\citet{Bowler2018}.
We recalculated the absolute magnitudes of galaxies using the new redshifts if available.
For object 279127 no spectroscopic redshift has been derived, despite spectral scans as part of REBELS, and so for ID279127 we instead use the photometric redshift throughout.

\subsection{Near-infrared imaging from VISTA and~\emph{HST}}
In this work we also incorporate near-infrared imaging of the sources.
The near-infrared probes the rest-frame UV emission of the galaxies at $z \simeq 7$, and hence is used to define the rest-frame UV slope, $\beta$.
We utilized both the ground-based UltraVISTA dataset~\citep{McCracken2013} in the $YJHK_s$ filters and single band $J_{140}$ data from~\emph{HST}/WFC3.
The~\emph{HST} data was obtained as follow-up and is described in detail in~\citet{Bowler2017}.
We used the most recent UltraVISTA data release 4 (DR4) imaging over the COSMOS field, which reaches $5\sigma$ depths of 26.2 ($Y$-band) to 25.3 ($K_s$-band) in a $1.8\,{\rm arcsec}$ diameter circular aperture in the ultra-deep region, where all six sources reside.
The imaging was used as provided with a pixel scale of $0.15\,{\rm arcsec}/{\rm pix}$ and point-spread functions (PSFs) were obtained from the software {\sc PSFex}~\citep{Bertin2013}.
For obtaining PSF homogenised imaging we created convolution kernels based on these output PSFs for each band and smoothed all images to the poorest resolution ($Y$-band).
All six sources were imaged with a single orbit of~\emph{HST}/WFC3, which reaches depths of $26.9$ in a  $0.6\,{\rm arcsec}$ circular aperture~\citep{Bowler2017}.
We re-reduced the $J_{\rm 140}$ data to take into account the change in astrometric solution between the UltraVISTA DR2 and the DR4 data (see next Subsection).
We combined the calibrated, flat-fielded, individual exposure files ({\sc flt}) from the archive using the {\sc Astrodrizzle} pipeline and a pixfrac of 0.8.
The~\emph{HST}/WFC3 imaging has a FWHM of 0.2 arcsec.

\begin{table}\caption{A summary of the Band 6 ALMA data utilized in this study to investigate the dust continuum properties of bright $z \simeq 7$ LBGs.
The upper part of the table presents the `high-resolution' dataset while the lower part presents the `low-resolution' dataset used to define if a source is detected.
Column 1 lists the ALMA observing cycle in which the data was taken.
Columns 2 and 3 give the integration time and the RMS found for that data, respectively.
The fourth column presents the beam size, which depends on whether a taper has been applied.
The taper scale is presented in Column 5, where `nat' corresponds to the naturally weighted data where no taper has been applied.
}\label{table:depths}
\begin{tabular}{ccccccr}
\hline
Cycle  & $t_{\rm int}$ & RMS & beam & taper & Sample \\
& $/{\rm min}$ & $/\mu{\rm Jy}/{\rm beam}$ & $/{\rm arcsec}$ & $/{\rm arcsec}$ & \\
\hline
6 &  119 & $13$ & $0.79 \times 0.62$ & nat & Excl. 65666\\
4  & 113 & 19 & $0.73 \times 0.66$ & 0.75 &  65666 \\

\hline
6 & 119 & $9$ & $1.36 \times 1.04$ & 0.50 & Excl. 65666 \\
3 & 10 & $27$ & $1.42 \times 1.08$ & nat &  65666 \\

\hline
\end{tabular}
\end{table}

\subsection{Astrometry}\label{sect:astrometry}

For our sample the dominant astrometric error in the ALMA data is due to the relatively low signal-to-noise ratio (SNR) of the detections, coupled with the size of the beam.
Assuming the beam size of our naturally weighted data (Table~\ref{table:depths}), with the approximation of the astrometric accuracy given by $\sigma = 0.6 \times {\rm FWHM}/{\rm SNR}$\footnote{ALMA Technical Handbook:\url{https://almascience.eso.org/documents-and-tools/cycle6/alma-technical-handbook}} leads to predicted positional errors of $[0.14, 0.08]$ arcsec for SNR $= [3, 5]$.
In the near-infrared, the UltraVISTA DR4 data has been astrometrically matched to the Gaia DR2 release\footnote{http://ultravista.org/release4/dr4\_release.pdf}, leading to expected errors that are sub-pixel and of the order of $\sim 0.05\,{\rm arcsec}$ with no offsets found between the different filters.
For the follow-up~\emph{HST}/WFC3 imaging, the absolute astrometric solution provided is poor due to the lack of precision in the pointing of the telescope.
Hence as part of the combination of the {\sc flt} files using {\sc Astrodrizzle}, we match the~\emph{HST}/WFC3 astrometry to Gaia DR2.
The reference epoch of the Gaia stars is set to 2016, which is the same year as the~\emph{HST} data was obtained, and we therefore do not consider any proper motion.
As the field of view of~\emph{HST} is very small, only 4-5 stars are matched directly between this data and the Gaia catalogue.
Hence in practice we match the astrometry to the UltraVISTA $J$-band image, which in turn has been tied to Gaia over a significantly larger area.
The resulting astrometric accuracy between the F140W imaging and UltraVISTA (and hence Gaia) is sub-pixel with observed offsets in the range 0.01-0.03 arcsec with an RMS scatter of around one~\emph{HST}/WFC3 pixel, or $0.06$ arcsec.
We also visually inspected the offsets to identify any trends that might be obscured when measuring the average, however we find no evidence for any simple astrometric transformation such as a rotation.
Finally, we also compared Very Large Array (VLA) positions of ALMA detected serendipitous sources in the field.
When matching to VLA catalogues that have a comparable beam size to our ALMA datasets we find no evidence for systematic errors, with an agreement within 0.05 arcsec.
The results of these checks indicate that our astrometric accuracy is very good, to within less than one~\emph{HST}/WFC3 pixel, and we use this information to give confidence to the observed offsets and colour gradients that we discuss in the next sections.

\section{Methods}\label{sect:methods}

In this section we describe the methods used to extract information from our ALMA,~\emph{HST} and VISTA datasets for our sample of six UV-bright $z \simeq 7$ star-forming galaxies.

\subsection{Measurement of the FIR dust continuum emission}
We used the Python Blob Detector and Source Finder ({\sc PyBDSF};~\citealp{Mohan2015}) software package for source extraction in the ALMA Band 6 data.
We used a detection threshold of $3\sigma$ and retained sources found within a radius of $1$ arcsec from the UltraVISTA $J+H$ centroid of each LBG in our sample.
We measured both the peak flux and the flux derived from a Gaussian fit to the data.
For the five previously undetected sources we define whether the galaxy is significantly detected or not in the ALMA data based `low-resolution' (tapered) dataset, to ensure we are sensitive to any extended emission.
From our single data point in the rest-frame FIR we estimated the total IR luminosity by assuming an SED shape and integrating over the rest-frame wavelength range of $8$--$1000\,\mu{\rm m}$.
To provide a simple comparison to previous studies, our fiducial measurement was made assuming a modified black-body (or greybody) fit which has two free parameters, the SED dust temperature ($T_{\rm dust}$) and the dust emissivity ($\beta_{\rm dust}$).
The lack of observational constraints on these two parameters at $z > 6$ provides one of the biggest uncertainties in the $L_{\rm IR}$ measurement.
We fix $\beta_{\rm dust} = 1.6$, corresponding to that found for IR luminous galaxies as presented in~\citet{Casey2012}.
Recent results have demonstrated a large spread in $T_{\rm dust} = 35$ to $80\,{\rm K}$ for individual sources~\citep{Knudsen2016, Bakx2020, Bakx2021, Sugahara2021, Sommovigo2021}.
We choose a dust temperature of $50\,{\rm K}$ in this work, which is consistent with that predicted by the evolution model of~\citet{Bethermin2015} and measurements of similarly luminous $z \simeq 5$ sources~\citep{Faisst2020}.
Choosing this value also allows a straightforward comparisons to other studies (e.g.~\citealp{Schouws2021}).
We note the deviation caused by an increase or decrease in dust temperature where appropriate.
In~\citet{Bowler2018} we found that assuming a different dust SED, such as one with a mid-IR component or an empirical template led to an increase in the derived $L_{\rm IR}$ by at most a factor of two.
The obscured SFR was derived from the $L_{\rm IR}$ using the calibration from~\citet{Kennicutt2012}, where we have converted to a~\citet{Chabrier2003} initial mass function (IMF).

\subsection{Search for \cii line emission}
The main goal of the new Cycle 6 ALMA observations we present here was to detect or obtain limits on the dust continuum emission of these high-redshifts sources.
However, as ALMA provides the full spectral data-cube prior to the creation of a continuum image, we are also able to search for the \cii emission line in the frequency range covered.
To search for potential emission lines we extracted both the peak flux and that in an elliptical aperture corresponding to the beam FWHM. 
We then binned the data to provide a frequency resolution of $40\,{\rm km}/{\rm s}$.
We tested that our method was able to detect the lines identified in~\citet{Smit2018a} from their archival data.
Estimating the expected position of the lines from the photometric redshift, only object ID238225 had an expected \cii frequency that overlapped with our Cycle 6 ALMA observations.
For this object and for the other sources we found no evidence for \cii emission at a significance of $>2\sigma$ in the cubes.
\citet{Schouws2021} and the REBELS collaboration (\citealp{Bouwens2021}; Schouws et al. in preparation) subsequently observed the same five sources as our Cycle 6 data.
Using spectral line scans they were able to detect \cii emission in four of these objects, confirming that this line was outside the spectral windows utilised in our observing program.
The spectroscopic redshifts derived from \cii for five of the six sources in our sample are shown in Table~\ref{table:ra}.

\subsection{Measurement of the rest-frame UV slope and \boldmath${M_{\rm UV}}$}
We extracted the global photometry for each galaxy from the UltraVISTA $YJHK_{s}$ imaging with {\sc SExtractor} using 1.8 arcsec diameter circular apertures.
Photometric errors were determined locally using blank aperture measurements and taking the median absolute deviation from the closest 200 apertures to each source.
We measured the rest-frame UV slope, $\beta$, from the best fitting SED model to the $YJH$ photometry. 
The $K_{s}$ band was excluded as it probes the near-UV wavelengths beyond the typical window used to estimate $\beta$ defined in~\citet{Calzetti1994} to be $\lambda_{\rm rest} = 1268$--$2580$\AA.
A power-law was fitted to the model in the standard~\citet{Calzetti1994} regions to avoid significant emission and absorption features.
For the SED fitting we used~\citet{Bruzual2003} models with exponentially declining star-formation histories, and a~\citet{Chabrier2003} initial mass function. 
A single metallicity of $1/5\,{\rm Z}_{\odot}$ was used.
The dust attenuation law was taken from~\citet{Calzetti2000} and the magnitude of the attenuation in the rest-frame $V$-band was allowed to vary in the range of $A_{\rm V} = 0$--$6$.
We fixed the redshift to the spectroscopic redshift if available, or the best-fit photometric redshift from the full fitting (including the optical bands;~\citealp{Bowler2014}).
We also experimented with including the $K_{s}$ band in the derivation of $\beta$, which led to consistent values within the errors, however we identified a bias to redder colours with this method.
Due to the detection of spatially separated emission between the near ($Y$, $J$) and far-UV ($K_{s}$) bands (see Section~\ref{sect:results}), we caution that this redder $\beta$ slope could be artificial and due to the combination of physically distinct regions within the galaxy.
The absolute UV magnitude, $M_{\rm UV}$, was determined by integrating the best-fitting SED model with a top-hat function centred on $\lambda_{\rm rest} = 1500\,$\AA~with a width of $100$\AA.
The observed unobscured SFR was estimated from $M_{\rm UV}$ using the~\citet{Madau1998} prescription, again corrected to a~\citet{Chabrier2003} IMF.

\subsection{Deconfusion of the ground-based VISTA data}
The LBGs we study in this work show highly extended and clumpy rest-frame UV emission~\citep{Bowler2017}, which leads to them being resolved not only with~\emph{HST} but also in the multi-band ground-based VISTA imaging.
We therefore attempted a deconfusion analysis to extract the fluxes of different rest-frame UV components of the galaxy in the UltraVISTA $YJHK_{s}$ filters.
The software package {\sc TPHOT} was used for this analysis~\citep{Merlin2015}, with the high-resolution image provided by the single band~\emph{HST}/WFC3 $J_{140}$ data.
As discussed in Section~\ref{sect:astrometry}, we have confidence that our astrometric accuracy between these low and high resolution data are consistent to within $\sim 0.05\,{\rm arcsec}$, which corresponds to one pixel in the WFC3 data and $1/3$ of a pixel in the UltraVISTA data.
To define the separate $J_{140}$ components we used {\sc SExtractor} with a minimum detection area of 10 pixels above $2\sigma$ significance.
This process ensured that the sources were split into at most two components.
Some sources show detail on smaller scales (e.g. ID169850), however these components are too faint and close to the peak of emission to extract robust multi-band constraints.
To measure the flux from each component we subtracted all neighbours using the {\sc TPHOT} modelling and then used a $1.8$ arcsec diameter circular aperture at the~\emph{HST} centroid.
The results using this aperture were found to be consistent with those obtained with a smaller aperture of $1.0$ arcsec in diameter.

\begin{figure}
\begin{center}
\includegraphics[width = 0.22\textwidth]{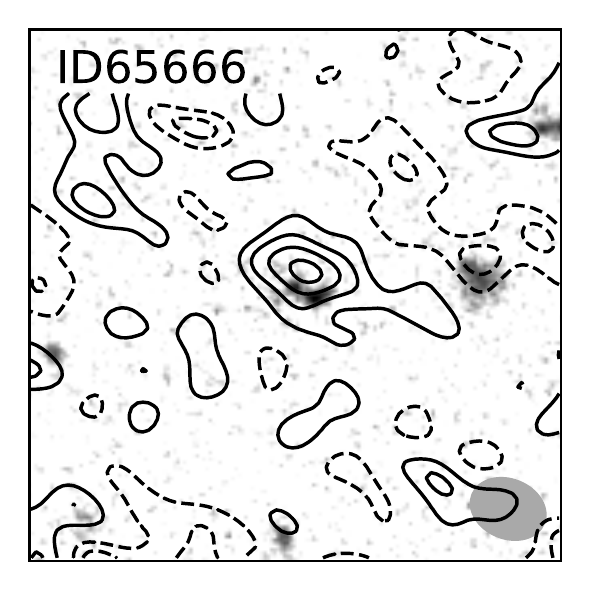}
\includegraphics[width = 0.22\textwidth]{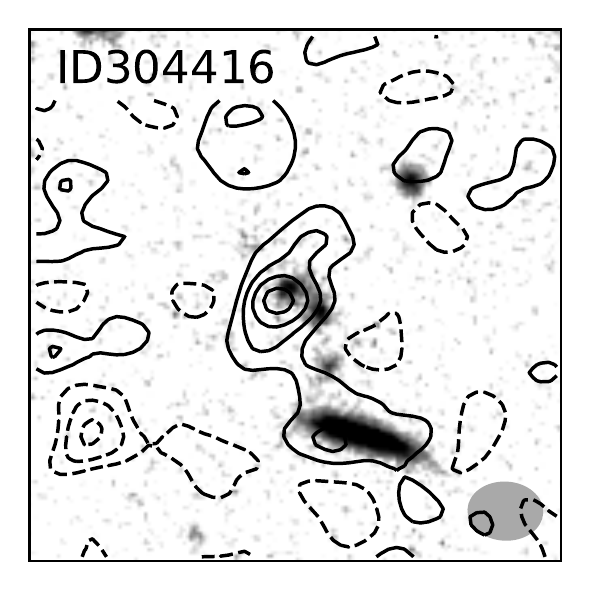}
\includegraphics[width = 0.22\textwidth]{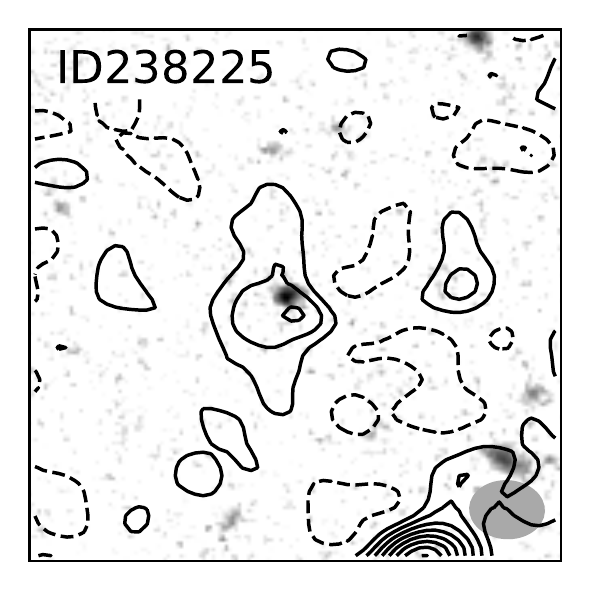}
\includegraphics[width = 0.22\textwidth]{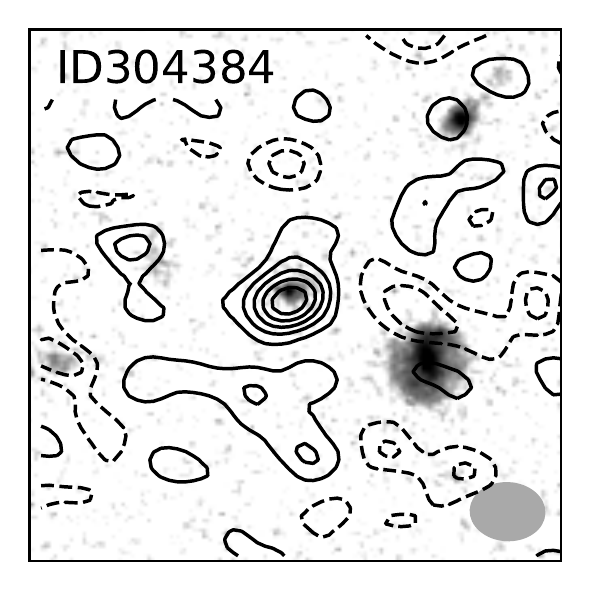}
\includegraphics[width = 0.22\textwidth]{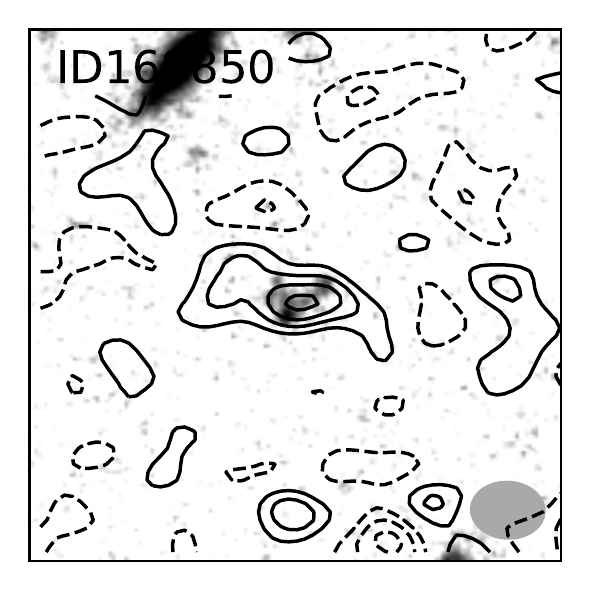}
\includegraphics[width = 0.22\textwidth]{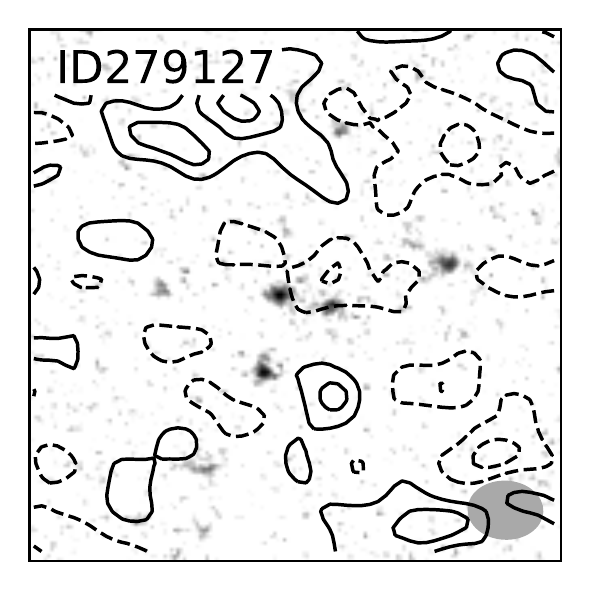}

\caption{The ALMA data probing the FIR dust continuum emission for our sample of $z \simeq 7$ UV-bright galaxies.
The contours, representing the ALMA Band 6 data, are shown at $1\sigma$ intervals with positive (negative) signal shown as the solid (dashed) lines.
Here we present the `low-resolution' tapered dataset, which has a resolution of 1.4 x 1.1 arcseconds.
The data for ID65666 is as shown in~\citet{Bowler2018}.
The background image is the~\emph{HST}/WFC3 data in the $J_{140}$-band, which probes the rest-frame UV emission at the redshift of these LBGs.
Each image is $10\,{\rm arcsec}$ on a side, with North to the top and East to the left.
We have scaled the~\emph{HST} $J_{140}$ band imaging from $[1, 5]\sigma$ in flux space.
The ID number is listed in the upper left of each stamp, and the beam is illustrated with the grey ellipse in the lower right hand corner.
}\label{fig:det}
\end{center}
\end{figure}

\section{Results}\label{sect:results}

In Fig.~\ref{fig:det} we present the result of our new, deeper, dust continuum observations in comparison to the~\emph{HST}/WFC3 imaging for the sample.
To determine the total flux and the proportion of sources with a detection in the FIR, we first focus on the `low-resolution' tapered data, to ensure we are measuring the full flux from the object regardless of morphology/spatial scales of the emission.
For the five previously undetected sources we now detect four objects at $> 3\sigma$ significance in our deeper Cycle 6 data.
Three of the sources are confidently detected at $>4 \sigma$, while one source, ID238225 (REBELS-30), is marginally detected at $\sim 3\sigma$.
The galaxy ID279127 (REBELS-31) is undetected with a $2\sigma$ upper limit of $< 21.5 \,\mu{\rm Jy}$.
The measured fluxes and derived properties are presented in Table~\ref{table:fluxes}.
The fluxes measured for the detected sources are on average $\sim 40\,\mu {\rm Jy}$ for the peak flux, or slightly higher at  $\sim 60\,\mu {\rm Jy}$ for the Gaussian fit.
This is somewhat lower than the estimate for these sources determined from the stacking analysis in~\citet{Bowler2018}, where we detected a signal of $\sim 100 \pm 50 \,\mu {\rm Jy}$, however it is consistent within the (large) errors.
In Fig.~\ref{fig:det} we also show the previous data for ID65666 as initially presented in~\citet{Bowler2018} as these Cycle 3 observations have a comparable SNR and resolution.
Combined with this previous data, we now detect five of the six brightest LBGs from the~\citet{Bowler2014} sample at $z \simeq 7$ in the dust continuum, allowing us to investigate the average dust obscured SFR without the need for stacking (see Section~\ref{sect:diss}).

\subsection{Dust continuum morphology and spatial offsets}

\begin{figure}
\begin{center}
\includegraphics[width = 0.22\textwidth]{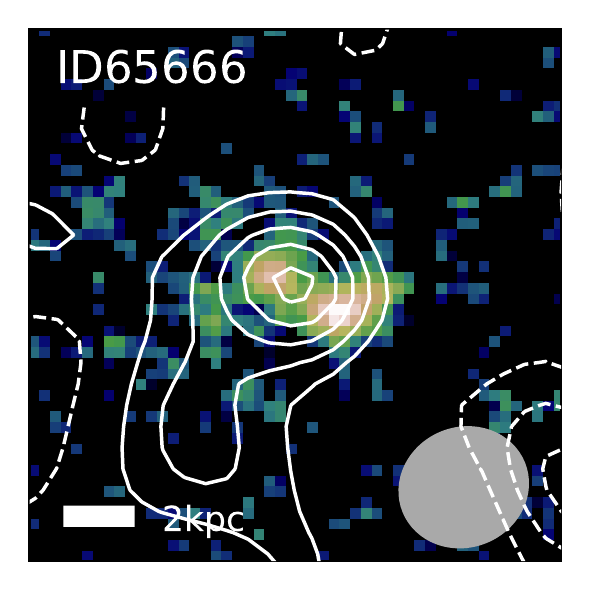}
\includegraphics[width = 0.22\textwidth]{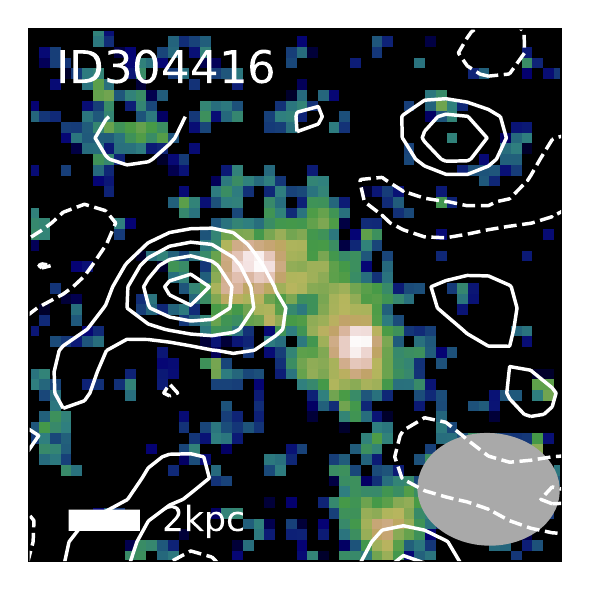}
\includegraphics[width = 0.22\textwidth]{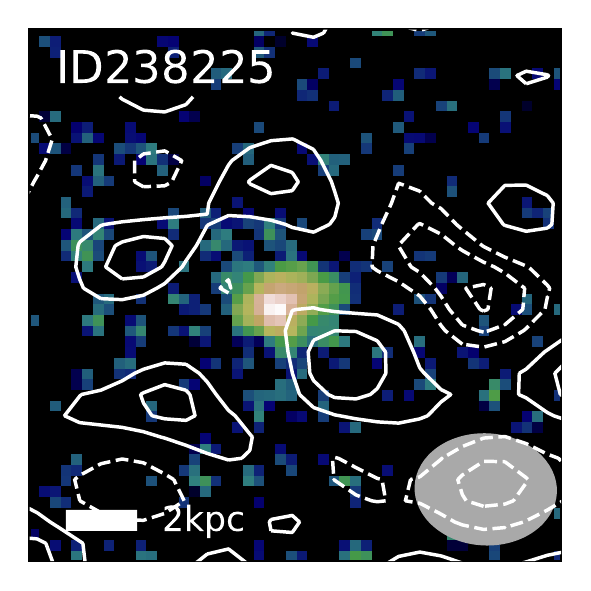}
\includegraphics[width = 0.22\textwidth]{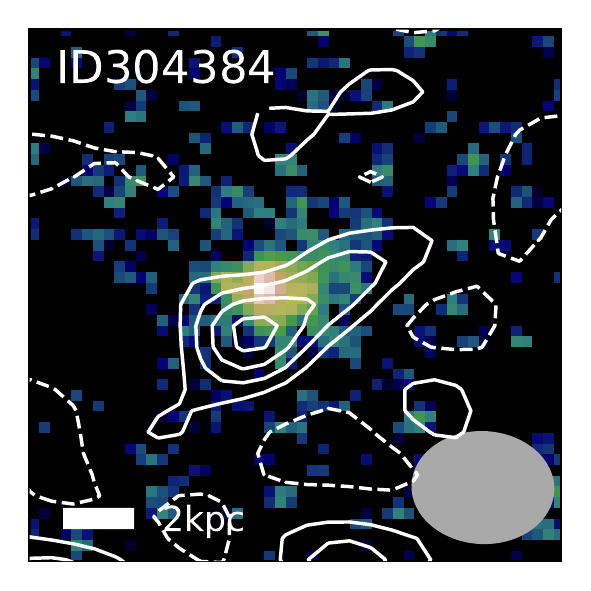}
\includegraphics[width = 0.22\textwidth]{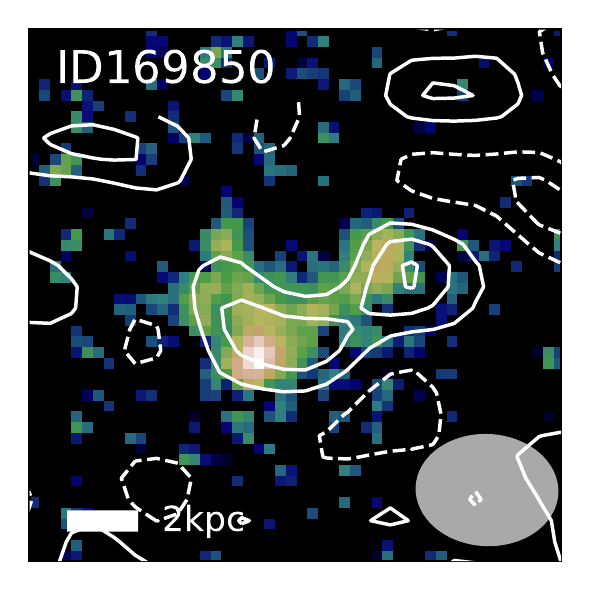}
\includegraphics[width = 0.22\textwidth]{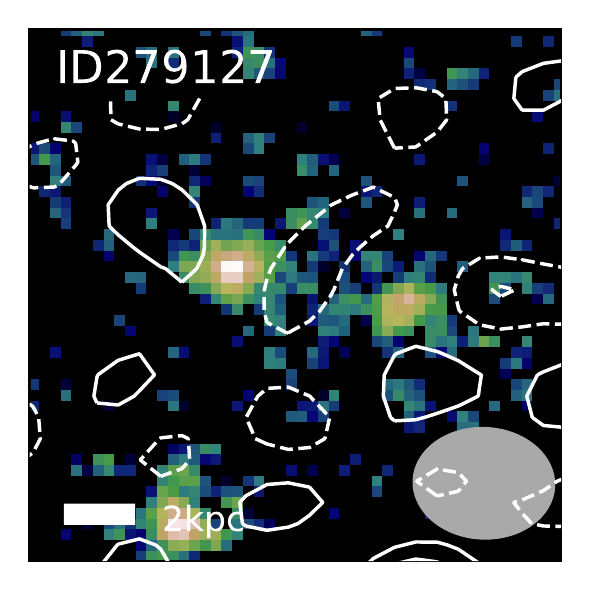}

\caption{The `high-resolution' ALMA dataset (white contours) shown in comparison to the~\emph{HST}/WFC3 imaging for our sample.
The stamps are $3 $\as~on a side and the background $J_{140}$ imaging has been scaled in magnitude space between the peak surface brightness and $26 \,{\rm mag}/{\rm arcsec}^2$. 
The beam and contour levels are as described in the caption of Fig.~\ref{fig:det}.
A scale bar in physical {\rm kpc} is shown in the lower left of each plot.
Three of the sources with significant dust detections show evidence for offsets between the centroid of the rest-frame UV emission and the ALMA contours (ID65666, ID304416 and ID304384).
}\label{fig:highres}
\end{center}
\end{figure}

The key advantage of our new ALMA observations is that they were obtained at relatively high spatial resolution, in comparison to previous studies of $z \simeq 7$ samples (e.g.~\citealp{Bowler2018, Schouws2021}).
In Fig.~\ref{fig:highres} we present the `high-resolution' dataset that includes the naturally weighted imaging from our Cycle 6 observing program, which has a beam FWHM $\simeq 0.7\,{\rm arcsec}$ (see Table~\ref{table:depths}).
We include the data from~\citet{Hashimoto2019} for source ID65666, tapered to the same resolution to allow a direct comparison\footnote{Note that this data does not have the same centroid as the~\citet{Bowler2018} detection shown in Fig.~\ref{fig:det}, and the potential explanations for this (including dust emission on different spatial scales) are discussed in~\citet{Hashimoto2019}.}.
We find a variety of dust morphologies within our sample of LBGs.
The two highest redshift detections (ID65666 and ID304416; top rows of Fig.~\ref{fig:det} and~\ref{fig:highres}) show compact emission that appears offset from the peak of the rest-frame UV emission as probed by the~\emph{HST}/WFC3 imaging.
In the case of ID65666 the dust continuum emission appears to sit between the two clumps seen in the rest-frame UV, however for object ID304416 the dust detection appears offset from both the observed~\emph{HST} clumps.
The emission from object ID304384 also appears offset from the~\emph{HST} centroid, which in this case is a single compact source (with a half-light radius of $r_{1/2} = 1.3 \pm 0.2\,{\rm kpc}$;~\citealp{Bowler2017}).
To quantify the significance of the offsets we considered both the possible astrometric systematic and random errors for the different datasets.
As discussed in Section~\ref{sect:astrometry}, we are confident in our astrometric solution for the near-infrared data to the order of one~\emph{HST} pixel.
The dominant positional error comes from the low SNR of the ALMA detections, hence we use the estimated Band 6 error, taking into account the angle of the separation relative to the position angle (PA) of the beam, to determine an offset significance for each source.
In comparison to the previous data presented in~\citet{Bowler2018}, where the error in the offset was $\simeq 0.2$\as, our new data leads to estimated positional errors of $<0.1$\as.
For ID304416 the derived offset between the ALMA detection and the closest rest-frame UV clump is $0.41 \pm 0.09$ arcsec, or a $4.6$ sigma detection excluding any error in the~\emph{HST} position.
If the light weighted centroid in the $J_{140}$-band is used instead, the separation increases to $0.62 \pm 0.09$ arcsec or $6.8\sigma$.
For ID304384 we find a smaller separation of $0.29 \pm 0.09$, corresponding to a significance of $3.1\sigma$.

In addition to apparently compact and potentially offset components we find evidence for extended rest-frame FIR emission as shown for object ID169850 and ID304384 (bottom-left and middle-right stamps respectively in Fig.~\ref{fig:det} and~\ref{fig:highres}).
Most strikingly, the flux for ID169850 appears to split into two components (of $2.6$ and $3.1\sigma$ significance) that are oriented in the same direction as the extended rest-frame UV emission.
Calculating the error in the ALMA positions for the two components as above, we find that it is coincident with the rest-frame UV emission to less than $0.1$ arcsec, or much less than 1$\sigma$ in this case.
Hence from our data we find that the FIR emission appears co-incident with the rest-frame UV clumps, although the positional error in the ALMA measurement is increased due to the low SNR of each component.
Object ID304384 (middle-right in Fig.~\ref{fig:det} and~\ref{fig:highres}) in also shows a possible extension with a PA of $\simeq 45$ degrees on the sky.
The detection of this extension is apparent in the $2$ and $3\sigma$ contours.
This source appears as a single compact component in the~\emph{HST}/WFC3 imaging, with no evidence for clumps or elongation in the rest-frame UV to the currently available depth and resolution.

The weakest ALMA detection in our data was for object ID238225, which was found to have a significance of $3.1\sigma$ in the tapered data shown in Fig~\ref{fig:det} (middle left).
In the naturally weighted imaging we do not recover a detection at $> 3\sigma$ significance.
The initial detection in the tapered data shows evidence for being extended both visually and in the observed difference between the peak flux and that obtained from a Gaussian fit.
Hence it is possible that we are unable to detect the flux at higher resolution because of the reduced surface brightness limit of this data (e.g. as found for \cii emission in~\citealp{Carniani2020}).
We find several peaks of $2\sigma$ significance within $1$\as~of the rest-frame UV position, however deeper data is required to understand any possible extended or multi-component dust continuum emission further.

\subsubsection{The size of the dust continuum emission}

We investigated measuring the sizes of the FIR emitting regions from our ALMA data.
As part of the source selection we obtained size estimates from {\sc PyBDSF}.
In the tapered data, three of the sources (ID65666, ID304416 and ID304384) show measured FWHM values that are consistent with being unresolved in our `high-resolution' dataset.
This is consistent with the fact that the derived peak fluxes are comparable to that derived from the Gaussian fit, although the large errors on the Gaussian parameters limit this comparison somewhat.
Objects ID238225 and ID169850 show evidence for extended emission as derived from the measured {\sc PyBDSF} sizes being larger than the beam size.
In addition they show the largest difference between the peak and Gaussian fit derived fluxes, which is consistent with some extended emission.
As discussed previously, object ID238225 shows no significant $> 3\sigma$ peaks within the naturally weighted data.
However, the difference between the peak and Gaussian flux in addition to a resolved size measurement both point to the source showing an extended region of dust continuum emission surrounding the observed rest-frame UV emission. 
For ID169850 the ALMA flux extension is confirmed in the naturally weighted data, where we observe two distinct clumps separated by $\sim 1\,{\rm arcsec}$. 
As a final test for whether the single component dust detections are resolved in our highest resolution images we manually subtracted the synthesised beam from our observations, and then searched for any significant residuals that would indicate an extended or multi-component source (following e.g.~\citealp{Hodge2016}).
The result of this test revealed no evidence for resolved or extended flux in sources ID304416 and ID65666 in the `high-resolution' dataset we present in Fig.~\ref{fig:highres}.
For object ID304284 we confirm the presence of the visually identified extended tail to be significant at $> 2\sigma$ once the beam is subtracted.
This is separated from the brightest peak of the ALMA emission by around $0.6\,{\rm arcsec}$.

\subsection{Resolved rest-frame UV colours}

\begin{figure*}
\begin{center}

\includegraphics[width = 0.49\textwidth, trim = 1.2cm 0.5cm 0.0 0.0cm]{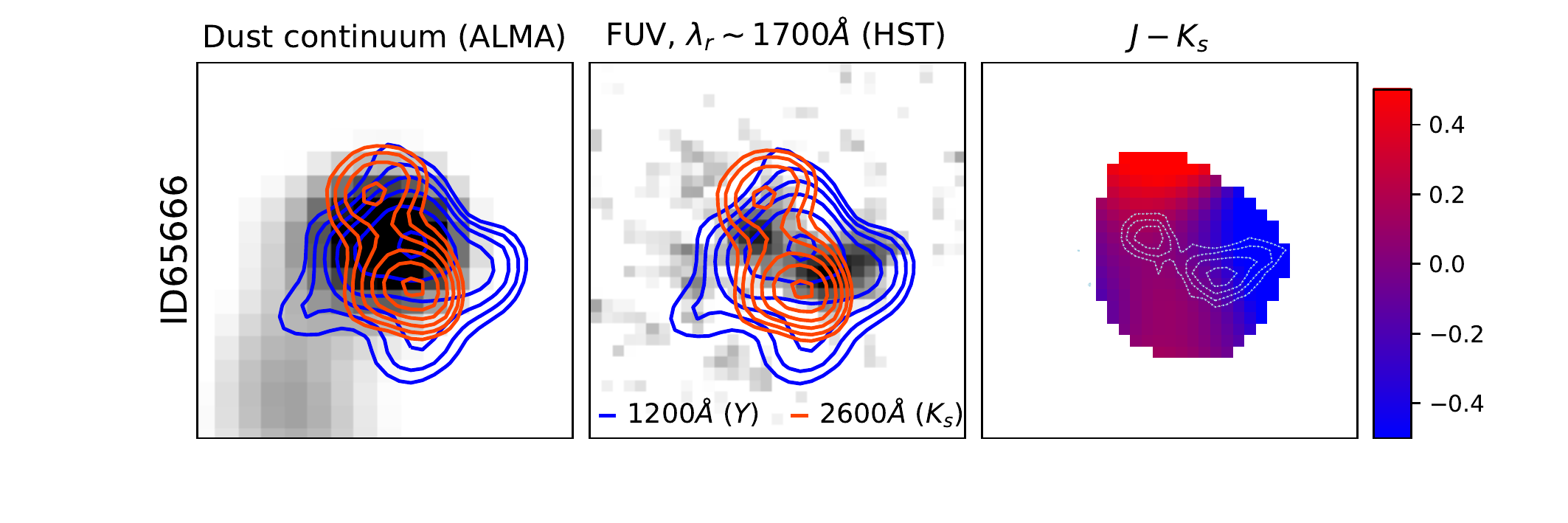}
\includegraphics[width = 0.49\textwidth, trim = 1.2cm 0.5cm 0.0cm 0.0cm ]{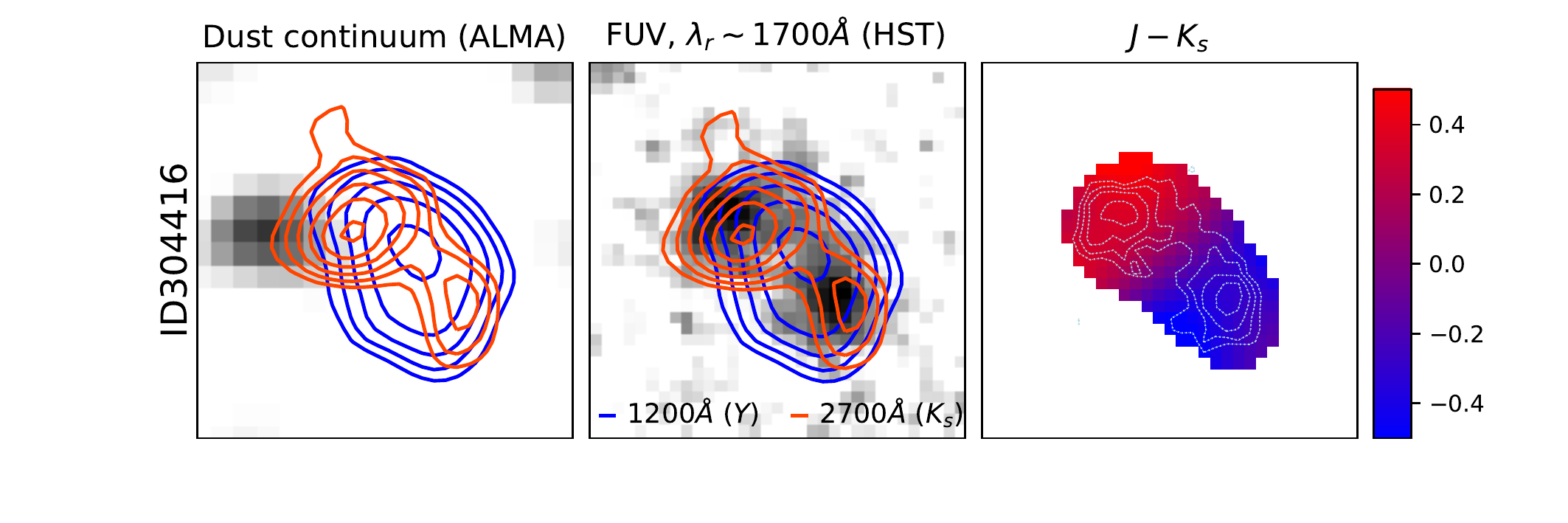}\\
\includegraphics[width = 0.49\textwidth, trim = 1.2cm 0.5cm 0 0.0cm]{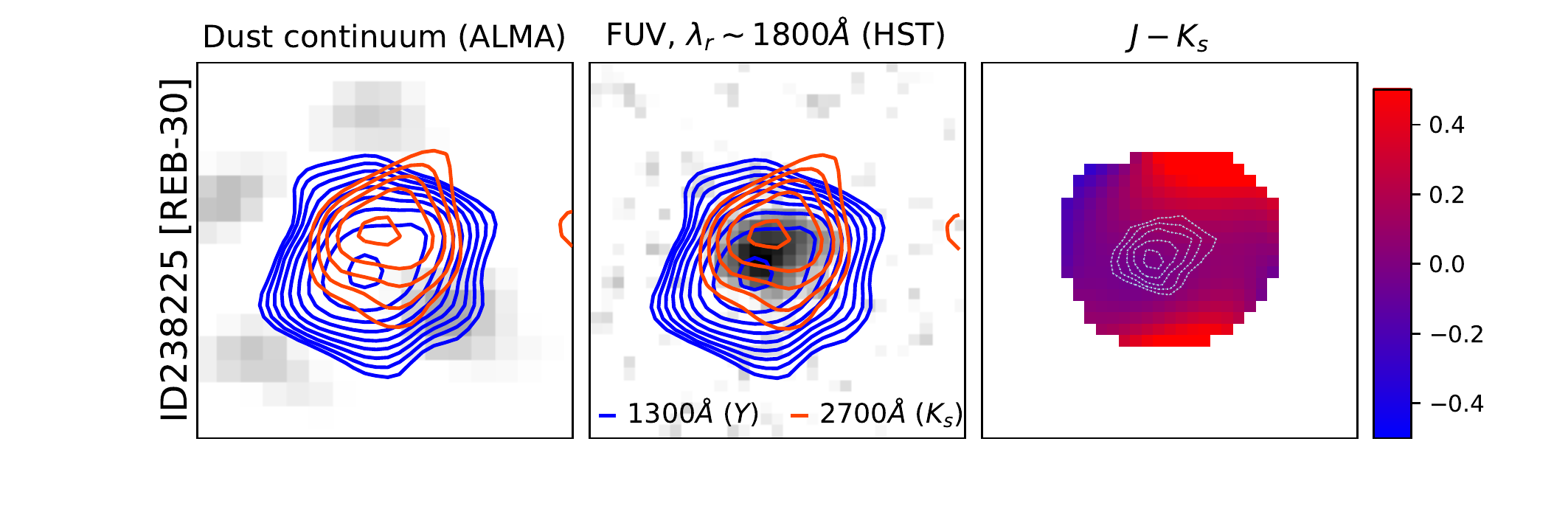}
\includegraphics[width = 0.49\textwidth, trim = 1.2cm 0.5cm 0 0cm]{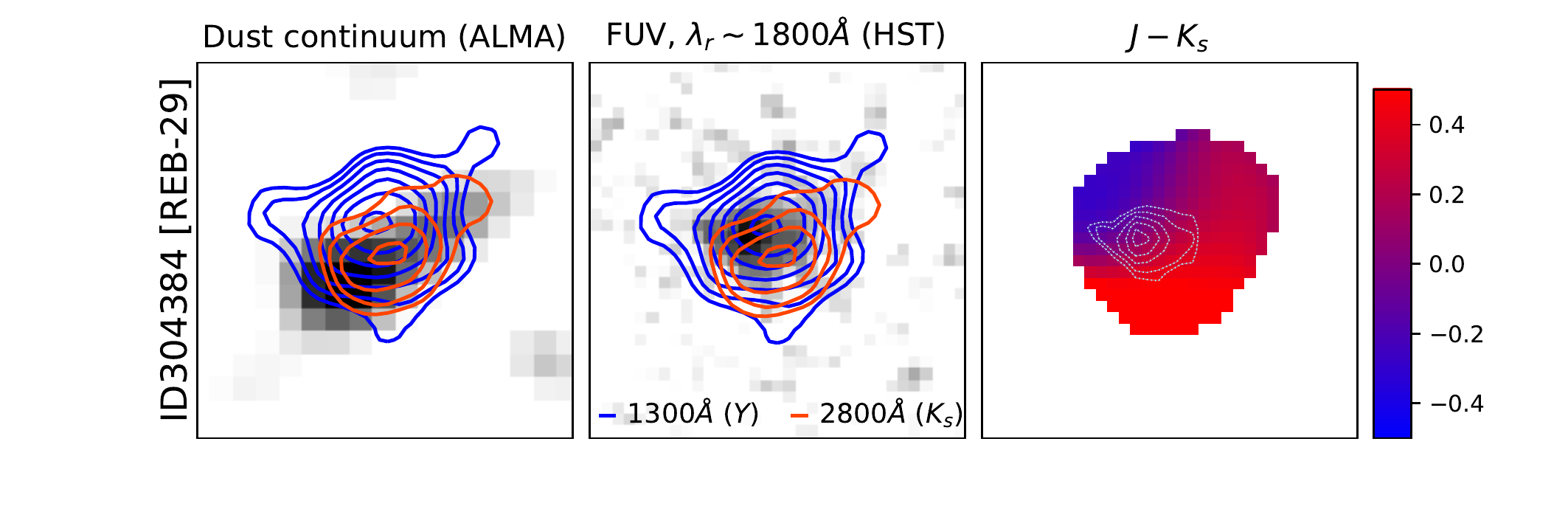}\\
\includegraphics[width = 0.49\textwidth, trim = 1.2cm 0.5cm 0 0.0cm]{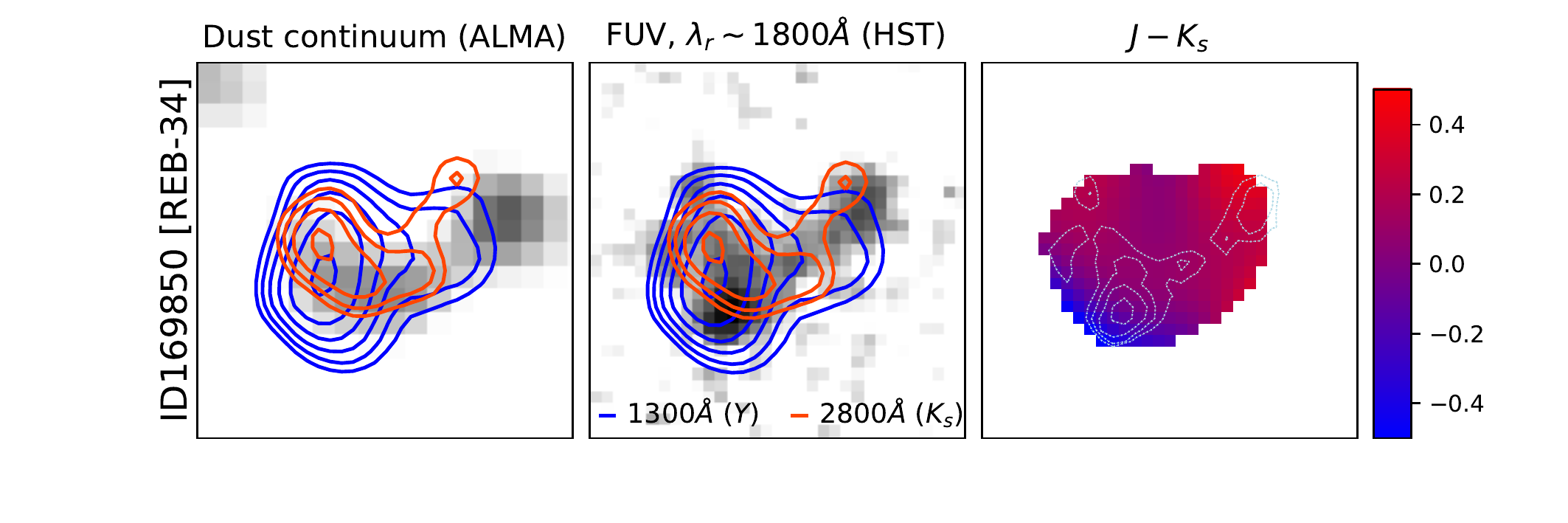}
\includegraphics[width = 0.49\textwidth, trim = 1.2cm 0.5cm 0 0.0cm]{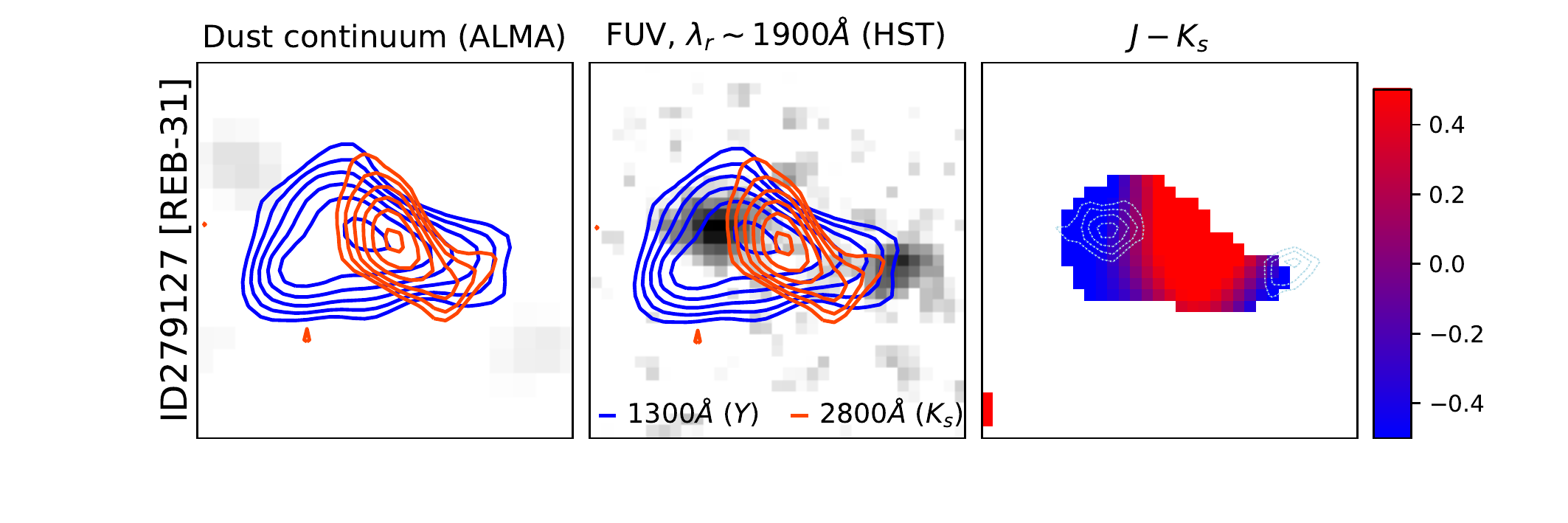}\\

\caption{
The resolved rest-frame UV colours for the bright $z \simeq 7$ galaxies studied in this work.
In the left and middle plot for each source we show the ground-based $Y$ and $K_s$-band imaging as the blue and red contours respectively.
The contours represent the surface brightness (SB) in steps of 0.15 mag between the peak SB and a level defined as $10\sigma$ per pixel.
The background image for the left plot shows the naturally weighted ALMA Band 6 data, scaled to between $[1, 5]\sigma$ per pixel, probing the rest-frame FIR dust continuum emission of the galaxies.
In the central plot the background image shows the rest-frame far-UV emission as measured by the~\emph{HST}/WFC3 data in the $J_{140}$-band.
The~\emph{HST} data is scaled as in Fig.~\ref{fig:highres}.
The title and caption of the central figure show the rest-frame wavelengths probed by each NIR filter.
The right-hand plot shows the resolved $J-K_s$ colour measured from the PSF homogenised ground-based imaging with the colour corresponding to the measured magnitude shown in the scale on the far right.
The blue contours in this plot show the $J_{140}$ data to aid in identifying which clumps correspond to different regions in the extended ground-based data.
}\label{fig:resolved}
\end{center}
\end{figure*}

As the~\emph{HST} imaging covering these sources is only available in a single band ($J_{140}$), the UltraVISTA $YJHK_s$ data provides the only insight we have into the rest-frame UV colour of the galaxies.
The fact that the bright $z \simeq 7$ galaxies we present in this work are extended on scales of $\ge 1$\as~in the rest-frame UV means that we can resolve them not only with the~\emph{HST}/WFC3 imaging, but also in the ground-based UltraVISTA data.
The seeing of the UltraVISTA data is in the range $\sim 0.7$--$0.9$\as~from the $K_{s}$ to the $Y$-band, allowing us to extract resolved colours for the sample.
In Fig.~\ref{fig:resolved} we present the multi-band UltraVISTA near-infrared data in comparison to the ALMA Band 6 data and the high resolution~\emph{HST}/WFC3 $J_{140}$ imaging.
We show the $Y$ and $K_{s}$ data as contours in surface brightness, scaled from the brightest pixel to a surface brightness limit corresponding to $10\sigma$ in that band.
The $K_{s}$ band has the highest spatial resolution and we compare this to the $Y$-band to provide the longest possible wavelength baseline with which to identify colour gradients.
Note that there could be Lyman-$\alpha$ emission situated in the $Y$-band, which could have a greater spatial extent in comparison to the stellar continuum (e.g.~\citealp{Steidel2011}).
We verified that the ground-based $J$-band contours and the $J_{140}$ imaging were co-spatial, as expected from our  astrometric checks.
In Fig.~\ref{fig:resolved} the most striking observation is that the $K_{s}$-band data, our highest resolution ground-based band, shows multiple components for several of the sources where such components are also seen in the~\emph{HST}/WFC3 $J_{140}$-band.
This observation clearly demonstrates the unique properties of the sample amongst known $z \simeq 7$ sources, where the brightness and extent means that they can be resolved into multiple components from the ground.
For ID169850 and ID304416 these components correspond to the $J_{140}$ clumps observed with~\emph{HST}/WFC3.
In the case of ID65666, the $K_{s}$ clumps appear somewhat offset and in a different orientation to the $J_{140}$ imaging.
For the $Y$ and $J$-band ground-based imaging we are limited by the poorer spatial resolution of these bands, which leads to the $J_{140}$ identified clumps being indistinguishable and merging into a single broad component.

From the qualitative comparison of the contours in the UltraVISTA data there is evidence for colours gradients within the sources, with the centroids of the $Y$, $J_{140}$ and $K_s$-band differing in all sources.
To assess the robustness of these colour gradients and to quantify the rest-frame UV slope of each clump, we proceed to analyse this data using two methods; first, by making a map of the $J-K_s$ colour and second, by undergoing a deconfusion analysis to extract the colours for separate~\emph{HST} components.
To create a colour map we used the PSF homogenised images and computed a $J-K_{s}$ colour (note that the two left-hand panels of Fig.~\ref{fig:resolved} are not PSF homogenised).
We excluded the $Y$ band for this measurement, as for three of the sources this band contains the Lyman-break, which could influence the measured continuum colour.
We only calculate the colour on pixels which have a detection in a $J + K_s$ stack at $2\sigma$ significance.
For all sources we observe colour gradients in the ground-based data.
Where there are multiple components, in our three brightest sources (ID65666, ID304416 and ID169850), the gradients appear to be associated with the different clumps as observed in the~\emph{HST} $J_{140}$ data.
Even in apparently single component sources we see a gradient with the centroid of the $Y$ and $K_s$-band images being offset from each other by 0.18 and 0.14 arcsec for ID304384 and ID238225 respectively, which is in most cases greater than any expected astrometric error.
The corresponding offsets between the $J$ and $K_s$-band were 0.15 and 0.05 arcsec for ID304384 and ID238225 respectively.

\begin{figure*}

\includegraphics[width = 0.32\textwidth, trim = 0 0.5cm 0 0 ]{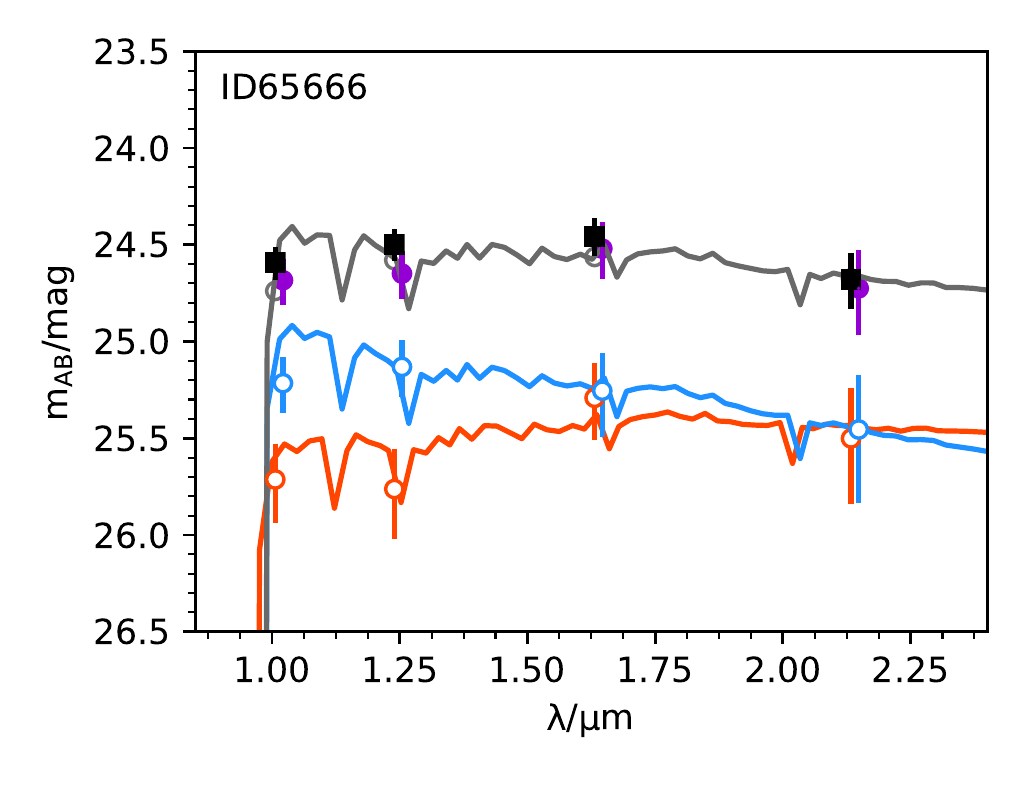}
\includegraphics[width = 0.32\textwidth, trim = 0 0.5cm 0 0 ]{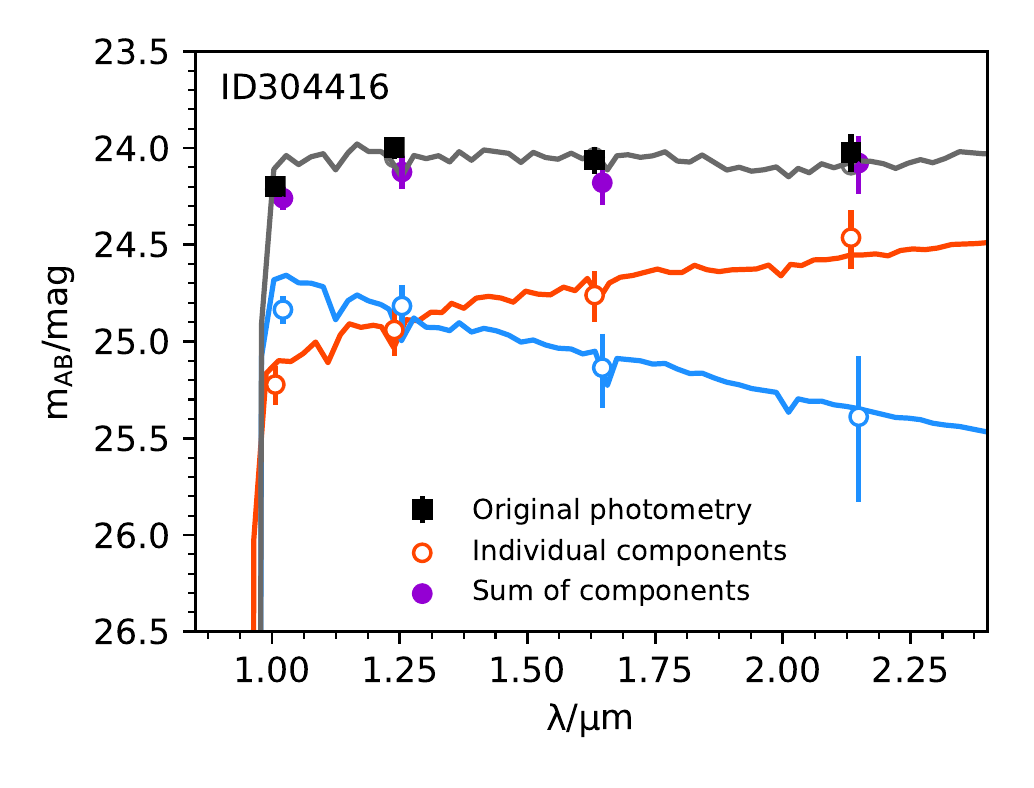}
\includegraphics[width = 0.32\textwidth, trim = 0 0.5cm 0 0 ]{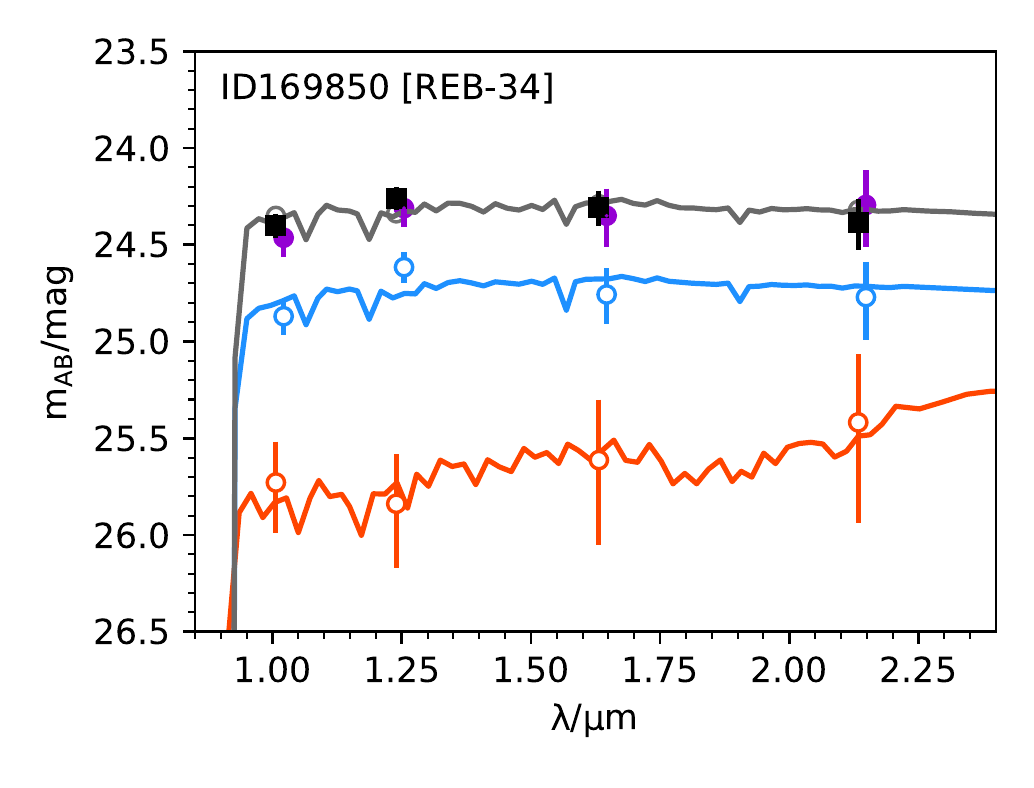}

\caption{The ground-based UltraVISTA $YJHK_s$ photometry extracted for each clump in the three multi-component sources in our sample.
The open circles show the results derived from {\sc TPHOT} for each component (two components for each galaxy).
The best-fitting SED template is shown as the solid lines for the bluer and redder components.
Reassuringly, the sum of the deconfused photometry (purple filled circles) agrees well with the total photometry measurement on the original data (black squares).
The grey line shows the best-fit SED model to this total photometry, with the grey open circles representing the synthesised photometry from this model.
}\label{fig:sed}
\end{figure*}

To quantify the colour gradients further we used a deconfusion analysis to extract the photometry in the different components of the multi-component sources.
We show the photometry extracted for the two brightest clumps in ID65666, ID304416 and ID169850 in Fig.~\ref{fig:sed}.
The results of this analysis confirm that the clumps show different colours as observed visually in Fig.~\ref{fig:resolved}.
For ID304416 and ID65666 the colours are clearly different, with one component being particularly blue ($J-K_s \simeq -0.4\,{\rm mag}$) and the other red ($J-K_s \simeq 0.4\,{\rm mag}$).
For ID169850 one of the components is significantly fainter than the other, and the colour difference is not as evident due to the larger errors on the photometry.
We verified that the sum of the two components in each source was consistent with the full 1.8\as~photometry, showing that we are not missing significant flux from these sources beyond these two main components.

How do these gradients in the rest-frame UV relate to the dust continuum emission?
In objects ID65666 and ID304416 (top row of Fig.~\ref{fig:resolved}) we see that the gradient across the multiple components is such that the reddest clump in the rest-frame UV is closest to the observed dust continuum emission.
This is also the case in ID169850, although we also see some ALMA flux at the position of the bluest component (as there are two faint ALMA components in this source).
Source ID304384 is the only source with a significant ALMA detection that is not resolved into multiple components in the~\emph{HST}/WFC3 data.
Interestingly, we observe a strong colour gradient in this source in the ground-based data, with the reddest side located on top of the elongated ALMA emission.
The galaxy ID238225 also shows a tentative gradient, however as we have no dust emission at $> 3\sigma$ and as the rest-frame UV emission is a single clump in $J_{140}$ we cannot comment further on the colour gradient with respect to the dust emission.
Finally, object ID279127 shows an unusual morphology in the ground-based data, with the $K_{s}$-band emission appearing between the two bluer clumps that are detected in the $J_{140}$ imaging.
Both of the two blue components are best-fit as high-redshift galaxies (with photometric redshifts of $z \simeq 6.6 \pm 0.1$), and hence it is possible that these components are the UV-bright clumps within an extended (and redder) disk.
Another possibility is that it is a strongly lensed system, where the central red source is a low redshift lensing galaxy.
Further follow-up to obtain spectroscopic redshifts of the clumps is required to understand this source further.

\begin{table*}
\caption{The measured flux and derived rest-frame UV and IR luminosities for the sample.
In the upper section of the table we present the global properties of the sources, while in the lower section we present the resolved measurements for the three sources (65666, 169850 and 304416) in which we observe multi-component emission.
In Column 1 we show the ID number (see Table~\ref{table:ra} for alternative IDs) followed by the absolute UV magnitude in Column 2.
The measured ALMA Band 6 flux is shown in Column 2 and 3 for the peak and Gaussian fit results respectively.
For the peak flux column we show the signal-to-noise ratio of any detection in brackets.
The IR luminosity is shown in Column 5, where we assume a greybody SED with $T_{\rm dust} = 50\,{\rm K}$, followed by the derived IRX in Column 6.
The SFR derived from the UV and the FIR via the scaling relations described in Section~\ref{sect:methods} are shown in Columns 7 and 8.
From these SFRs we derive an obscured SFR fraction which is shown in Column 9.
Finally in Column 10 we present the measured rest-frame UV slope, $\beta$.
}\label{table:fluxes}
\begin{tabular}{cccccccccc}
\hline
ID & $M_{\rm UV}$ & $f_{\rm peak}$ & $f_{\rm Gauss}$ & $L_{\rm IR}$ & ${\rm log}_{10}({\rm IRX})$ & ${\rm SFR}_{\rm UV}$ & ${\rm SFR}_{\rm FIR}$ & $f_{\rm obs}$ & $\beta$ \\
& $/{\rm mag}$ & $/\mu{\rm Jy}$ & $/\mu{\rm Jy}$ & $/10^{11}$ \Lsun& & \sfrunit & \sfrunit & & \\
\hline
65666 & $ -22.52_{-0.09}^{+0.08}$ & $ 124.5 \pm 28.0 \, (4.4)$ & $ 156.4 \pm 53.4 $ & $ 6.4 \pm 2.2$ & $ \phantom{-}0.47_{-0.21}^{+0.16}$ & $ 31_{-2}^{+3}$ & $ 89_{-30}^{+30}$ & $ 0.74 \pm 0.26$ & $ -2.21_{-0.28}^{+0.18}$  \\[1ex]
304416 & $ -23.15_{-0.06}^{+0.05}$ & $ 42.2 \pm 8.6 \, (4.9)$ & $ 48.1 \pm 19.6 $ & $ 2.0 \pm 0.8$ & $ -0.28_{-0.25}^{+0.17}$ & $ 55_{-3}^{+3}$ & $ 28_{-1
1}^{+11}$ & $ 0.34 \pm 0.14$ & $ -2.32_{-0.07}^{+0.21}$  \\[1ex]
238225 & $ -22.35_{-0.11}^{+0.10}$ & $ 32.3 \pm 9.5 \, (3.4)$ & $ 81.7 \pm 48.6 $ & $ 3.6 \pm 2.1$ & $ \phantom{-}0.29_{-0.43}^{+0.25}$ & $ 26_{-2}^{+3}$ & $ 49_{-29
}^{+29}$ & $ 0.65 \pm 0.39$ & $ -2.01_{-0.30}^{+0.22}$  \\[1ex]
304384 & $ -22.31_{-0.11}^{+0.10}$ & $ 52.4 \pm 8.9 \, (5.9)$ & $ 50.0 \pm 15.0 $ & $ 2.1 \pm 0.6$ & $ \phantom{-}0.08_{-0.20}^{+0.16}$ & $ 25_{-2}^{+3}$ & $ 29_{-9}
^{+9}$ & $ 0.54 \pm 0.17$ & $ -1.90_{-0.19}^{+0.23}$ \\[1ex]
169850 & $ -22.78_{-0.06}^{+0.06}$ & $ 38.1 \pm 8.5 \, (4.5)$ & $ 53.2 \pm 32.5 $ & $ 2.3 \pm 1.4$ & $ -0.07_{-0.43}^{+0.23}$ & $ 39_{-2}^{+2}$ & $ 33_{-2
0}^{+20}$ & $ 0.45 \pm 0.28$ & $ -1.95_{-0.15}^{+0.15}$ \\[1ex]
279127 & $ -22.57_{-0.11}^{+0.10}$ & $ < 21.5\,(0.5)$ & $--$ & $ < 0.9 $ & $ < -0.38 $ & $ 32_{-3}^{+3}$ & $ < 13 $ & $ < 0.29 $ & $ -2.35_{-0.47}^{+0.23}
$ \\[1ex]

\hline
65666 E & $ -21.43_{-0.26}^{+0.21}$ & $ 90.9 \pm 17.8 \, (5.1)$ & $--$ & $ 3.7 \pm 0.7$ & $ 0.67_{-0.18}^{+0.19}$ & $ 11_{-2}^{+3}$ & $ 52_{-10}^{+10}$ & $ 0.82 \pm 0.24$ & $ -1.78_{-0.43}^{+0.28}$  \\[1ex]
65666 W& $ -21.87_{-0.16}^{+0.14}$ & $ < 53.8$ & $--$ & $ < 2.2 $ & $ < 0.27 $ & $ 17_{-2}^{+3}$ & $ < 31 $ & $ < 0.64 $ & $ -2.53_{-0.29}^{+0.23}$  \\[1ex]
304416 E& $ < -20.98$ & $ 58.6 \pm 13.2 \, (4.4)$ & $--$ & $ 2.5 \pm 0.6$ & $  0.68_{-0.11}^{+0.09} $ & $ < 7 $ & $ 34_{-8}^{+8}$ & $ > 0.82 $ & --  \\[1ex]
304416 NE& $ -22.03_{-0.13}^{+0.12}$ &  $ < 23.6$ & $--$ & $ < 1.0 $ & $< -0.16 $ & $ 20_{-2}^{+3}$ & $ < 13 $ & $ < 0.41 $ & $ -1.38_{-0.33}^{+0.19}$  \\[1ex]
304416 SW& $ -22.14_{-0.12}^{+0.11}$ & $ < 23.6$ & $--$ & $ < 1.0 $ & $ < -0.20 $ & $ 22_{-2}^{+3}$ & $ < 13 $ & $ < 0.38 $ & $-2.81^{+0.27}_{-0.19}$\\[1ex]
169850 W& $ -21.04_{-0.33}^{+0.25}$ & $ 42.6 \pm 13.6 \, (3.1)$ & $--$ & $ 1.9 \pm 0.6$ & $ 0.53_{-0.28}^{+0.27}$ & $ 8_{-2}^{+3}$ & $ 26_{-8}^{+8}$ & $ 0.77 \pm 0.
33$ & $ -1.74_{-0.53}^{+0.44}$  \\[1ex]
169850 E& $ -22.12_{-0.08}^{+0.08}$ & $ 34.7 \pm 13.6 (2.6)$ & $--$ & $ 1.5 \pm 0.6$ & $ 0.01_{-0.25}^{+0.18}$ & $ 21_{-1}^{+2}$ & $ 21_{-8}^{+8}$ & $ 0.50 \pm 0.20$ & $
-1.88_{-0.23}^{+0.23}$ \\[1ex]

\hline

\end{tabular}
\end{table*}

\section{Discussion}\label{sect:diss}

In this work we have presented deep ALMA observations targeting the rest-frame FIR of six UV bright $z \simeq 7 $ LBGs.
We detect five of these galaxies in the FIR at $> 3\sigma$ confidence (one source was previously detected in~\citealp{Bowler2018}), allowing us to provide some of the most complete constraints on the dust properties of star-forming galaxies within the EoR.
Furthermore, because of the relatively high spatial resolution of this data we can compare the rest-frame FIR sizes and morphologies of these sources to that observed in the rest-frame UV.
As shown in Table~\ref{table:fluxes}, the FIR luminosities of these galaxies are around $L_{\rm IR} = 2$--$6 \times 10^{11}\,$\Lsun~(assuming a greybody SED with $T_{\rm dust} = 50\,{\rm K}$).
While the sample represents some of the most highly star-forming LBGs known at $z > 6$, their measured FIR luminosities and total SFRs are at least an order of magnitude below those found for the highly dust obscured SMGs (which have $L_{\rm IR} > 10^{12}\,$\Lsun; e.g.~\citealp{Hodge2016}).
Thus they represent an interesting transition between SMGs and the typical high-redshift LBG population, which has previously shown evidence for being dust poor (e.g.~\citealt{Capak2015, Bouwens2016}).
In comparison to previous studies of SMGs and lower redshift LBGs the galaxies in our sample have lower stellar masses, with ${\rm log}_{\rm 10}(M_{\star}/{\rm M}_{\odot}) \simeq 9-9.5$~\citep{Bowler2018}.
Combined with the derived total SFRs (Table~\ref{table:fluxes}), we derived high sSFRs ($\sim 100\,/{\rm Gyr}$) for the galaxies, consistent with being in a starburst phase (e.g. as compared to $L \simeq L^*$ galaxies with sSFR $\sim 10\,/{\rm Gyr}$;~\citealp{Stark2013}).
The clumpy morphology observed in the rest-frame UV for the sources in our sample, coupled with the observed compact dust emission suggests, that this elevated sSFR may have been triggered by a merger.
Similar arguments have been made for SMGs (e.g.~\citealp{Chen2015}), although there is disagreement over whether SMGs are starbursts or simply high-mass main-sequence galaxies~\citep{Michalowski2017}.
Our sample therefore provides a unique insight into the distribution of young unobscured stars and dust within galaxies with stellar masses of ${\rm log}_{\rm 10}(M_{\star}/{\rm M}_{\odot}) \simeq 9-9.5$ at high redshift with which to compare to other dusty star-forming galaxies and fainter/less massive LBGs at the same epoch.

\subsection{IRX-\boldmath$\beta$ relation for bright \textbf{$z \simeq 7$} LBGs}
\begin{figure}

\includegraphics[width = 0.45\textwidth]{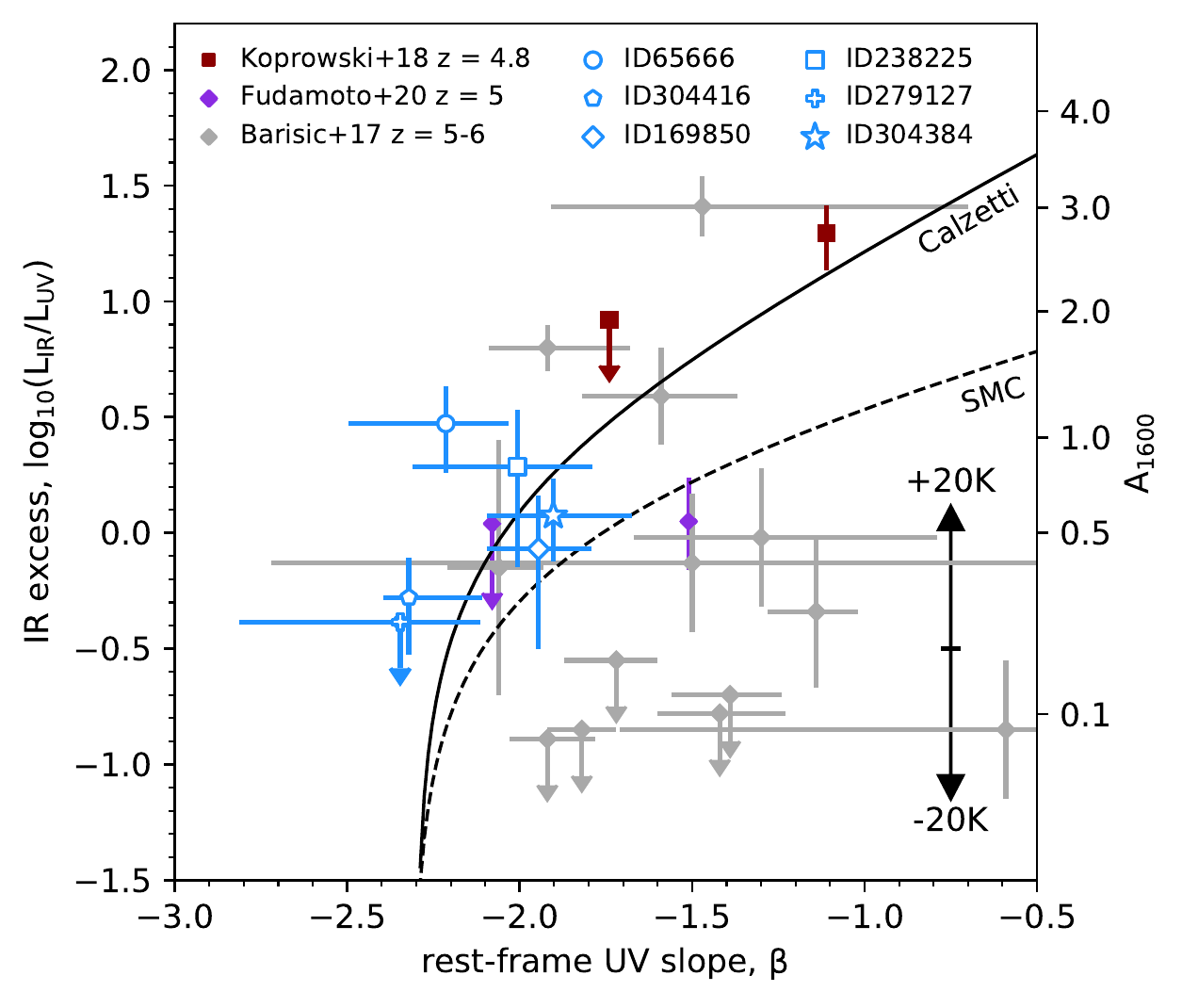}

\caption{The \irxb~values derived for the six bright $z \simeq 7$ LBGs in our sample in comparison to previous results.
We plot each galaxy individually using open symbols in blue.
The Meurer relation for Calzetti-like dust is shown as the black solid line, while the relation derived from assuming the SMC extinction curve is shown as the dashed line.
Previous results at $ > 4$ are shown from~\citet{Koprowski2018},~\citet{Fudamoto2020} and~\citet{Barisic2017} with red squares, purple diamonds and grey diamonds, respectively.
We assume a dust temperature of $50\,{\rm K}$.
The mean impact of changing the dust temperature by $\pm 20\,{\rm K}$ is illustrated by the arrows in the lower right.
The right-hand axis illustrates the expected dust attenuation at $1600$\AA~corresponding to that IRX value, according to the relation found for $z \sim 3$ sources in~\citet{McLure2018} (which is comparable to the Meurer relation).
}\label{fig:irx}
\end{figure}

By combining our rest-frame UV colour and luminosity with our new FIR detections we can now plot our sample on the \irxb~relation without the need for stacking (as was performed previously for the same galaxies in~\citealp{Bowler2018}).
In Fig.~\ref{fig:irx} we present the global \irxb~relation for our sample, where we take a single value of the IRX and $\beta$ to be representative of the full galaxy.
We compare our results with the local relation for starburst galaxies, as defined by a Calzetti-like dust attenuation law~\citep{Calzetti2000}, in addition to that predicted using a steeper dust law based on the SMC extinction curve.
What is striking from Fig.~\ref{fig:irx} is that the full sample is very blue, with $\beta \lesssim -2$.
From the extrapolation of the colour magnitude relation found at $z \simeq 7$ from~\citet{Bouwens2014beta} we would expect these bright sources with $M_{\rm UV} < -22.4$ to have $\beta \simeq -1.5$.
As we discuss further in Section~\ref{sect:dissUV}, we do find components of the galaxies that are redder (as well as bluer) than $\beta = -2$, however the global $\beta$ measurement provides an average colour that is weighted in favour of the brightest (and bluest) rest-frame UV clumps.
Our measured global \irxb~results agree with the Calzetti-like relation, under the assumption that a dust SED with $T_{\rm dust} = 50\,{\rm K}$ is reasonable.
ID65666 appears somewhat above the relation, with a bluer than expected rest-frame UV slope given the measured IRX.
This region of the diagram has previously been found to be populated by IR-luminous dusty star-forming galaxies (DSFGs) at $z \lesssim 3 $ (e.g.~\citealp{Casey2014a, Clark2015, Dunne2018}), where it has been postulated that an inhomogeneous dust distribution leads to bluer colours than expected.
ID65666 has a clumpy and irregular rest-frame UV and FIR morphology (also see~\citealp{Hashimoto2019}) in parallel with these DSFGs.
Furthermore~\citet{Casey2014a} found that the deviation blue-ward from the canonical \irxb~relation was a function of $L_{\rm IR}$, and we note that ID65666 is the most IR luminous source in our sample, as well as showing the highest total SFR and sSFR~\citep{Bowler2018}.
Hence the offset above the Calzetti-like \irxb~relation of this source could indicate that the same trend with $L_{\rm IR}$ found for DSFGs is present in $z \simeq 7$ LBGs, indicating that extreme bursts of star-formation can cause the observed offset in~\irxb, however clearly larger samples are required to confirm this observation.

We do not observe the same spread in $\beta$-slopes to red values as found by~\citet{Capak2015} and~\citet{Barisic2017}.
This could be due to differing selection functions, as the latter studies were based on spectroscopically confirmed LBG candidates.
We find good agreement with an extrapolation of the $z \simeq 5$ results of~\citet{Koprowski2018} and~\citet{Fudamoto2020}, although we note that these studies diverge at redder $\beta$ values not probed by our sample.
In comparison to the recent study of~\citet{Schouws2021}, our results for individual sources agree within the errors on the \irxb~plane, however they conclude that their results are most consistent with an attenuation curve between an SMC and Calzetti-like shape.
The IRX values we measure are also a proxy for the fraction of obscured star-formation.  
In Table~\ref{table:fluxes} we present the unobscured ${\rm SFR}_{\rm UV}$ derived from the measured $M_{\rm UV}$, and the obscured ${\rm SFR}_{\rm IR}$ derived from our computed $L_{\rm IR}$.
The obscured fractions for our sample where we have a detection are in the range of $0.35$--$0.75$, with an average of $f_{\rm obs} = 0.54$.
These values, which are consistent with those found for similar luminosity sources by~\citet{Schouws2021}, demonstrate that dust obscured star-formation is significant in the brightest LBGs at $z \simeq 7$.

We note that the derived $L_{\rm IR}$ values are highly dependent on the assumed dust temperature.
In Fig.~\ref{fig:irx} we show the effect of changing the dust temperature with a vertical arrow to illustrate changes of the order of $\Delta T_{\rm dust} = \pm 20\,{\rm K}$, which is comparable to the expected range from previous studies.
In the case of a higher dust temperature, our results would sit somewhat above the Calzetti-like relation, in a region of the \irxb~diagram that is predicted theoretically to be due to clumpy, irregular, obscured sources where the global $\beta$ value is dominated by unobscured regions of the source (e.g.~\citealp{Popping2017, Liang2021}).
In the converse situation, where the dust temperature is closer to $30\,{\rm K}$, we would still find consistency with the Calzetti-like relation due to the blue rest-frame UV slopes of our sources.
As is evident in this Figure, we do not have the dynamic range to set strong constraints on the underlying dust attenuation laws, as the typically assumed parameterisations of SMC and Calzetti-like dust asymptote to the same value corresponding to an intrinsic $\beta$-slope around where our measured galaxies lie.
Larger samples of sources at $z \simeq 7$, such as those being observed with the REBELS survey, are clearly required to fully constrain the \irxb~relationship within the EoR.

\subsection{The morphology of dust continuum emission}\label{sect:dissize}
Within our sample we find a diversity of dust continuum morphologies, with compact emission offset from the rest-frame UV in addition to extended and multi-component emission (Fig.~\ref{fig:highres}).
Similar features have been found within previous studies of sources at $z \simeq 5.5$ from the sample of~\citet{Capak2015, Barisic2017} and the larger ALPINE survey~\citep{LeFevre2020}.
\citet{Schouws2021} also find evidence for compact and offset dust emission in their brightest FIR source Y-003 at $z = 7.3$.
These results indicate that complex dust continuum morphologies may be common at high redshifts. 
To understand these results further it is instructive to look at studies of $z \simeq 1$--$3$ sources that have gained comparable or better spatial resolution in the rest-frame FIR.
As an example,~\citet{Rujopakarn2019} observed three `typical' main-sequence star-forming galaxies with $M_{\star} \sim 10^{10.5-11.0}\,{\rm M}_{\odot}$ at $z = 3$ with deep 200pc resolution ALMA follow-up.
In their sample the clumpy UV emission was observed to be offset by around $2$--$10\,{\rm kpc}$ from the bulk of the star-formation pinpointed by the FIR detection.
Offset UV and FIR emission and redder components hosting the bulk of the SFR of the galaxy appear to be common features found in samples of both lower-redshift LBGs and SMGs at high redshifts (e.g.~\citealp{Rivera2018a, Chen2017, Hodge2019, Chen2020, Cheng2020, Cochrane2021}).
Our results for UV bright LBGs at $z \simeq 7$ therefore fit into a common morphological class found in luminous star-forming galaxies.
Whether the observed clumps and irregularity is evidence for major mergers in these sources is still an open question.
Deeper observations to redder wavelengths have revealed disks underlying the clumpy rest-frame UV emission, which is not representative of the full galaxy extent/morphology (e.g.~\citealp{Wuyts2012, Targett2013}).
Using a kinematic analysis of the FIR \cii line allows come constraints on whether the components are merging (e.g.~\citealp{Hashimoto2019}) or in a rotating disk (e.g.~\citealp{Smit2018a}), however it is still challenging with current data to distinguish between these two scenarios in typical LBGs.
Observations with the~\emph{James Webb Space Telescope (JWST)} will be crucial to understand the context of the rest-frame UV and FIR components observed in this study, with the NIRCam/NIRSpec instruments able to probe the rest-frame optical wavelengths (and emission lines) for the first time in a resolved sense.
Such observations will be able to ascertain the morphology of the stellar mass distribution, as well as producing a map of the dust attenuation from spatially resolved measurements of the Balmer decrement.

\subsubsection{Compact dust emission in comparison to the UV?}
Another common observation in lower redshift studies is that the FIR dust continuum emission is more compact than the rest-frame UV/optical, typically by a factor of 2 (e.g.~\citealp{Fujimoto2017, Chen2017, Tadaki2020}).
The sizes of the FIR clumps in these studies appear to occupy a broad range from $r_{1/2} \simeq 0.5$--$5\,{\rm kpc}$ (see~\citealp{Tadaki2020, Cheng2020}).
Although we cannot robustly measure the sizes of our sources in the FIR, we can comment on the fact that they appear unresolved in our data and compare this to what is expected from these lower-redshift studies.
As the beam of our naturally weighted ALMA observations is approximately FWHM $= 0.7\,{\rm arcsec}$ (corresponding to a physical size of $\sim 4\,{\rm kpc}$), we would expect to resolve the FIR detections if the galaxies were as large as $r_{1/2} \sim 5\,{\rm kpc}$.
The sizes of our sample as derived from the rest-frame UV were presented in~\citet{Bowler2017}, where the average half-light radius of a stacked profile of bright sources was found to be $r_{1/2} = 2.30^{+0.21}_{-0.76}\,{\rm kpc}$.
This stacked size does not fully account for the large separation in different components found in the images, which are of the order of $5\,{\rm kpc}$.
If the rest-frame UV clumps are embedded in a larger disk, then the clump separation would be a more reasonable size estimate.
Hence for the brightest, multi-component sources in our sample (ID65666, ID304416) we can tentatively conclude that our FIR sizes are more compact than the observed UV sizes, consistent with that found at lower redshift.
This is not the case for all galaxies in our sample however, as two sources show evidence for extended/multi-component FIR emission of a comparable extent to the rest-frame UV size.
It may be the case that in lower mass, less extreme, star-forming galaxies the dust emission is more homogeneously distributed throughout the galaxy.
Larger samples of galaxies observed with high spatial resolution, for example like that upcoming with the ALMA large program `CRISTAL' (2021.1.00280.L; PI: Rodrigo Herrera-Camus), will further reveal the diversity of dust morphologies relative to the rest-frame UV emission within high-redshift LBGs. 

\subsection{Rest-frame UV colour gradients and resolving IRX-\boldmath$\beta$}\label{sect:dissUV}
As part of our analysis of the morphology and colours of our sample we have identified colour gradients in the rest-frame UV light seen across the spatial extent of the bright $z\simeq 7$ galaxies.
In the case where we observe bright and well separated clumps in the~\emph{HST} imaging, the colour gradients are very clear (e.g. ID304416), however we also find evidence for gradients even in the sources which appear to only have one UV component under~\emph{HST} resolution.
The observed gradients appear to be aligned such that the reddest regions are located closest to the FIR dust continuum detection.
This observation strongly suggests that the colour gradients are due to increasing dust obscuration, rather than other effects that could cause differential reddening such as a gradient in stellar age or metallicity (although these parameters will also be correlated with the existence of dust).
These results show that the UV brightest component is not necessarily the position with the highest SFR.
As an example, for ID304416 it is in-fact the `HST-dark' East clump that shows the highest total SFR of the three well defined components in this galaxy.

\begin{figure}

\includegraphics[width = 0.45\textwidth]{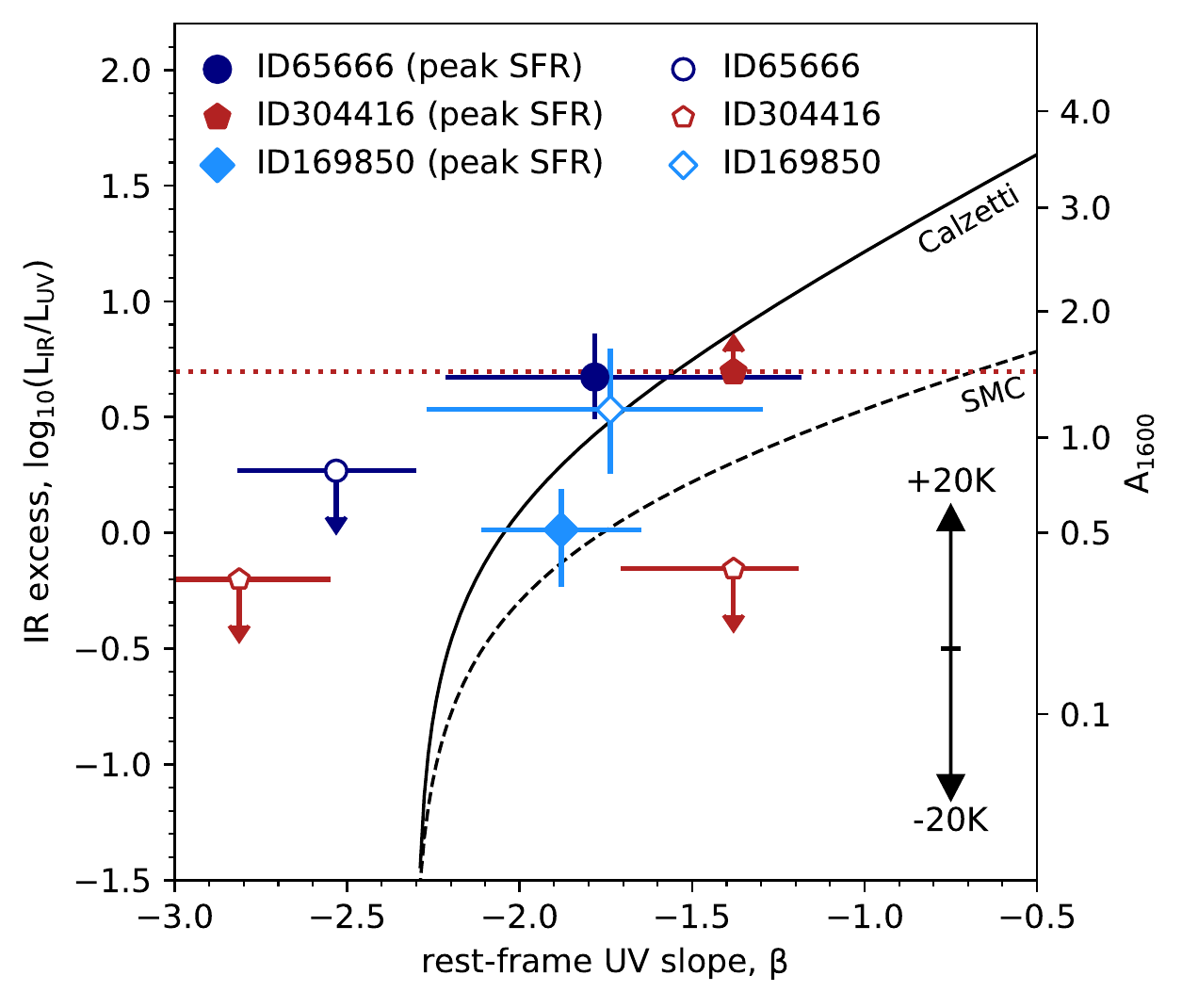}

\caption{The resolved \irxb~values derived for the individual deconfused components of the three sources in our sample that show multiple-components in the rest-frame UV.
Object ID304416 has been split into three clumps, one of which is undetected in the NIR and hence we plot the unconstrained $\beta$ of this source as a dotted horizontal line.
The other objects split into two components.
We have highlighted the region with the highest total SFR as a solid symbol.
Further details of the figure are as described in the caption of Fig.~\ref{fig:irx}.
}\label{fig:irxres}
\end{figure}

In Table~\ref{table:fluxes} we present the rest-frame UV colours, IRX and SFRs for each component of the three multi-component sources.
A deconfusion analysis of these objects demonstrated clear colour gradients between the observed clumps in the UV (see Fig.~\ref{fig:sed}).
To investigate the fraction of obscured star-formation across the resolved galaxies we associate each FIR dust detection with the closest rest-frame UV component if the separation was less than $3\sigma$, where $\sigma$ was taken as the error in the ALMA position calculated from the SNR and beam size (see Section~\ref{sect:astrometry}).
For ID65666 we associated the dust detection with the faintest UV clump, while for ID304416 the FIR detection is sufficiently offset from the two clear UV components to not be associated with either.
We therefore list the obscured component, ID304416 E, separately, with upper limits on the $M_{\rm UV}$ determined from aperture photometry on the $J_{140}$ image at that position.
For ID169850 we associate the two faint ALMA detections with the two main clumps in the $J_{140}$ image.
With these resolved measurements we are able to plot the \irxb~relation for each individual clump.
Note that typically the \irxb~diagram is created for global galaxy photometry, not for individual regions, and hence we caution that these results are not necessarily comparable to lower-redshift compendiums.
We nevertheless present our resolved \irxb~values in Fig.~\ref{fig:irxres} to provide a visual representation of the different rest-frame UV colours and obscured star-formation fractions across the face of each of the galaxies.
For the two highest redshift sources in our sample we find a striking colour gradient between the two well separated rest-frame UV components.
This is also visible in the colour maps shown in Fig.~\ref{fig:resolved}.
For galaxy ID304416 we measure $\Delta \beta = 1.43 \pm 0.35$, while for ID65666 we measure $0.75 \pm 0.44$.
A similar colour gradient has been found in the $z_{\rm spec} = 7.03$ source `DP7' as presented in~\citet{Pelliccia2021}.
They found a $\Delta \beta = 1.3$ between the two~\emph{HST} components of a $0.3\,L^*$ source, indicating that dichromatic sources such as those found in this work are present in both bright and faint LBGs at $z = 7$.
We find evidence for particularly blue components in both of these sources, showing $\beta \lesssim -2.5$, indicating dust-free components with young ages and low-metallicities (e.g~\citealp{Cullen2017}).
For galaxy ID169850, the other object for which we can obtained resolved colours, we find that the two main components have consistent colours with $\Delta \beta = 0.14 \pm 0.54$.
This source also shows extended FIR emission that is co-incident with the~\emph{HST}/WFC3 emission.
Combined with the lack of a strong rest-frame UV colour gradient, this observation could suggest that the galaxy has a more uniform dust distribution than the other resolved sources in our sample.

In Fig.~\ref{fig:irxres} we also highlight the component of each galaxy that has the highest total SFR.
We find that for $2/3$ of these multi-component sources, the peak SFR in the galaxy corresponds to the reddest component, with the peak in the SFR of object ID304416 being undetected in our~\emph{HST}/WFC3 imaging.
For this ID304416 E component we cannot measure a rest-frame UV slope with current data, hence we plot the $\beta$ value as a horizontal line in this figure.
The obscured fractions calculated for the regions of peak SFR in galaxies ID65666 and ID304416 are $f_{\rm obs} \gtrsim 0.80$, demonstrating clearly that a view of these LBGs based entirely on rest-frame UV data is insufficient.
These obscured components are comparable to the two `\emph{HST}-dark' galaxies found by~\citet{Fudamoto2021}, which were found to have an estimated $f_{\rm obs} > 0.75 (0.97)$ at $z = 7.35 (6.68)$.
In fact, the most obscured component found by~\citet{Fudamoto2021}, galaxy REBELS-29-2, was found only $40\,$ physical ${\rm kpc}$ from one of our targets (ID304384/REBELS-29).
The dust continuum detection for REBELS-29-2 is recovered in our dataset.
These results show that the unobscured rest-frame UV emission found within the target LBG (that motivated the observations of both this program and REBELS) is actually only revealing a small part of the total star-formation in the region, which is predominantly obscured.
The resolved measurements are roughly consistent with the Calzetti-like \irxb~relation found for local starburst galaxies.
For one component of ID304416 we measure red colours but no appreciable FIR detection, leading to a derived IRX that is well below even the SMC-like extinction curve.
Spatially this component sits between the FIR bright, obscured component and a very blue unobscured component, and it is challenging with current data to determine how much UV and FIR flux to associate with each clump.
Theoretical works have attempted to provide a physical interpretation of why galaxies lie in different regions of the \irxb~diagram.
The diagram is a summation of several interconnected factors, such as the dust geometry, stellar age and intrinsic $\beta$-slope (e.g.~\citealp{Popping2017, Liang2021}).
We speculate that within young, irregular galaxies within the EoR that a multitude of these astrophysical effects may be occurring within single sources (for example an evolved dust-free component next to a clump of young dusty star-formation), and as we start to collect spatially resolved data for $z \gtrsim 6$ objects we will begin to populate many regions of the \irxb~relation even within the same galaxy (see also~\citealp{Koprowski2016}).

\subsection{Future observations to redder wavelengths}

The data we have presented for the rest-frame UV and FIR emission from our sample of bright $z \simeq 7$ galaxies represents some of the first spatially resolved observations of galaxies within the EoR.
They represent an exciting taste of future spatially resolved observations, such as higher resolution data with ALMA and deep~\emph{JWST} imaging and spectroscopy.
What do we expect to see in deeper observations that go to redder wavelengths (e.g. with the~\emph{JWST} NIRCam)?
Building upon previous results at $z \sim 3$~\citep{Targett2013, Cheng2020, Cochrane2021} and theoretical works such as the resolved predictions of~\citet{Cochrane2019, Popping2021}, it is likely that the observed rest-frame UV morphology is not representative of the galaxy as a whole.
This is confirmed by our observations that show that the peak of the star-formation in our LBGs are often associated with the faintest, and reddest, components found in the rest-frame UV.
We would expect redder wavelength observations to probe older stellar populations, and be less affected by dust attenuation, and thus potentially be smoother than the observed rest-frame UV emission.
In comparison to previous resolved datasets for lower redshift LBGs and SMGs, it is likely that longer wavelength observations will progressively fill in the full underlying galaxy, of which the unobscured clumps observed in the UV are only pinpointing small disconnected regions~\citep{Cheng2020}.
The predicted morphology to redder wavelengths changes if the sources presented in this work are merging galaxies, where we might expect the underlying stellar mass distribution to break into multiple-components in a similar way to the rest-frame UV.

\section{Conclusions}\label{sect:conc}
In this work we present deep ($t_{\rm int} \simeq 120$ min per source) ALMA Band 6 imaging of six luminous $z \simeq 7$ galaxies with $M_{\rm UV} < -22.4$.
In total we detect five of the six sources at $> 3\sigma$ significance, providing one of the most complete samples of ALMA-detected LBGs at $z \ge 5$.
The IR luminosities of the sources were found to be $L_{\rm IR} = 2$--$6\,\times\,10^{11}\,{\rm L}_{\odot}$ (assuming a greybody dust SED with $T_{\rm dust} = 50\,{\rm K}$), placing them at least an order magnitude below the IR luminosities of SMGs.
We find that the fraction of obscured star-formation in our sample is in the range $f_{\rm obs} = 0.35$--$0.75$, indicating that on average more than half of the total SFR within bright LBGs is obscured.
Despite this observation, the sample shows particularly blue rest-frame UV colours, with $\beta \le -2$.
When placed on the \irxb~diagram, our sources are in good agreement with the relation found for local starburst galaxies with Calzetti-like dust, with no evidence for a steeper dust attenuation law such as the SMC-like relation found in other studies at $z \ge 5$.

Using our relatively high resolution (FWHM $\simeq 0.7\,{\rm arcsec}$) data we are able to investigate the dust morphologies of our sample and compare this to the resolved rest-frame UV light.
We find both compact dust emission that is offset from the observed rest-frame UV clumps, in addition to extended dust emission, either in the form of two similar components separated by $\sim 1\,$arcsec or in the detection of an elongated tail.
Through an analysis of the multi-band ground-based NIR data available for the sample we identify rest-frame UV colour gradients in the majority of the galaxies.
For the three most extended and multi-component sources we extracted resolved colours in the rest-frame UV using a deconfusion analysis.
We found a dramatic change in UV-slope across these sources with up to $\Delta \beta \simeq 0.7$--$1.4$.
The redder regions of each source were found to be located closest to the peak of FIR emission as probed by ALMA, indicating that the galaxy is being progressively obscured in that direction.
Our results show that the brightest UV clumps in $z \simeq 7$ galaxies are not necessarily representative of the true centroid of star-formation even in apparently blue LBGs.
We find that the fainter, redder, regions are typically hosting the peak of the SFR, with these components showing high obscured fractions of $f_{\rm obs} \ge 0.80$.
The spatially resolved data we have presented for the rest-frame UV and FIR emission for bright $z \simeq 7$ galaxies clearly demonstrates that there is a diversity of dust morphologies present in these sources, and that a homogeneous dust distribution cannot be assumed.
These observations give an exciting preview of upcoming resolved observations with~\emph{JWST} and ALMA that will further reveal the detailed structure of galaxies within the EoR.

\section*{Acknowledgements}
We thank the referee, Tom Bakx, for comments that improved the paper.
RB acknowledges support from an STFC Ernest Rutherford Fellowship [grant number ST/T003596/1].
This paper makes use of the following ALMA data: ADS/JAO.ALMA\#2018.1.00933.S, ADS/JAO.ALMA\#2016.1.00954.S and ADS/JAO.ALMA\#2015.1.00540.S. ALMA is a partnership of ESO (representing its member states), NSF (USA) and NINS (Japan), together with NRC (Canada), MOST and ASIAA (Taiwan), and KASI (Republic of Korea), in cooperation with the Republic of Chile. The Joint ALMA Observatory is operated by ESO, AUI/NRAO and NAOJ.
Based on data products from observations made with ESO Telescopes at the La Silla Paranal Observatory under ESO programme ID 179.A-2005 and on data products produced by CALET and the Cambridge Astronomy Survey Unit on behalf of the UltraVISTA consortium.  This research is based on observations made with the NASA/ESA Hubble Space Telescope obtained from the Space Telescope Science Institute, which is operated by the Association of Universities for Research in Astronomy, Inc., under NASA contract NAS 5--26555. These observations are associated with program \#13793.

\section{Data Availability}
The ALMA data presented in this paper are publicly available via the ALMA archive\footnote{\url{https://almascience.nrao.edu/aq/}}.
The program ID numbers are presented in Section~\ref{sect:data}.
The UltraVISTA $YJHK_{s}$ DR4 data are available via the European Southern Observatory Archive\footnote{\url{http://archive.eso.org/scienceportal/home?data_collection=UltraVISTA&publ_date=2019-03-11}} and the~\emph{HST}/WFC3 $J_{140}$ imaging can be accessed as part of the Mikulski Archive for Space Telescopes\footnote{\url{https://mast.stsci.edu/}}.




\bibliographystyle{mnras}
\bibliography{/Users/bowler/Documents/bibtex/library_abbrv} 

\begin{thebibliography}{}
\makeatletter
\relax
\def\mn@urlcharsother{\let\do\@makeother \do\$\do\&\do\#\do\^\do\_\do\%\do\~}
\def\mn@doi{\begingroup\mn@urlcharsother \@ifnextchar [ {\mn@doi@}
  {\mn@doi@[]}}
\def\mn@doi@[#1]#2{\def\@tempa{#1}\ifx\@tempa\@empty \href
  {http://dx.doi.org/#2} {doi:#2}\else \href {http://dx.doi.org/#2} {#1}\fi
  \endgroup}
\def\mn@eprint#1#2{\mn@eprint@#1:#2::\@nil}
\def\mn@eprint@arXiv#1{\href {http://arxiv.org/abs/#1} {{\tt arXiv:#1}}}
\def\mn@eprint@dblp#1{\href {http://dblp.uni-trier.de/rec/bibtex/#1.xml}
  {dblp:#1}}
\def\mn@eprint@#1:#2:#3:#4\@nil{\def\@tempa {#1}\def\@tempb {#2}\def\@tempc
  {#3}\ifx \@tempc \@empty \let \@tempc \@tempb \let \@tempb \@tempa \fi \ifx
  \@tempb \@empty \def\@tempb {arXiv}\fi \@ifundefined
  {mn@eprint@\@tempb}{\@tempb:\@tempc}{\expandafter \expandafter \csname
  mn@eprint@\@tempb\endcsname \expandafter{\@tempc}}}

\bibitem[\protect\citeauthoryear{Bakx et~al.,}{Bakx et~al.}{2020}]{Bakx2020}
Bakx T. J. L.~C.,  et~al., 2020, \mn@doi [MNRAS] {10.1093/mnras/staa509}, 493,
  4294

\bibitem[\protect\citeauthoryear{Bakx et~al.,}{Bakx et~al.}{2021}]{Bakx2021}
Bakx T. J. L.~C.,  et~al., 2021, MNRAS Letters, 508, 58

\bibitem[\protect\citeauthoryear{Bari{\v{s}}i{\'{c}}
  et~al.,}{Bari{\v{s}}i{\'{c}} et~al.}{2017}]{Barisic2017}
Bari{\v{s}}i{\'{c}} I.,  et~al., 2017, \mn@doi [ApJ]
  {10.3847/1538-4357/aa7eda}, 845, 41

\bibitem[\protect\citeauthoryear{Bertin}{Bertin}{2013}]{Bertin2013}
Bertin E.,  2013, ASCL, p. 1301.001

\bibitem[\protect\citeauthoryear{B{\'{e}}thermin et~al.,}{B{\'{e}}thermin
  et~al.}{2015}]{Bethermin2015}
B{\'{e}}thermin M.,  et~al., 2015, \mn@doi [A{\&}A]
  {10.1051/0004-6361/201425031}, 573, A113

\bibitem[\protect\citeauthoryear{Bethermin et~al.,}{Bethermin
  et~al.}{2020}]{Bethermin2020}
Bethermin M.,  et~al., 2020, A{\&}A, 643, A2

\bibitem[\protect\citeauthoryear{Bouwens et~al.,}{Bouwens
  et~al.}{2014}]{Bouwens2014beta}
Bouwens R.~J.,  et~al., 2014, \mn@doi [Astrophysical Journal]
  {10.1088/0004-637X/793/2/115}, 793, 115

\bibitem[\protect\citeauthoryear{Bouwens et~al.,}{Bouwens
  et~al.}{2016a}]{Bouwens2016}
Bouwens R.~J.,  et~al., 2016a, \mn@doi [ApJ] {10.3847/0004-637X/830/2/67}, 830,
  67

\bibitem[\protect\citeauthoryear{Bouwens et~al.,}{Bouwens
  et~al.}{2016b}]{Bouwens2016b}
Bouwens R.~J.,  et~al., 2016b, \mn@doi [ApJ] {10.3847/1538-4357/833/1/72}, 833,
  72

\bibitem[\protect\citeauthoryear{Bouwens et~al.,}{Bouwens
  et~al.}{2020}]{Bouwens2020}
Bouwens R.,  et~al., 2020, \mn@doi [ApJ] {10.3847/1538-4357/abb830}, 902, 112

\bibitem[\protect\citeauthoryear{Bouwens et~al.,}{Bouwens
  et~al.}{2021}]{Bouwens2021}
Bouwens R.~J.,  et~al., 2021, preprint (arXiv:2106.13719)

\bibitem[\protect\citeauthoryear{Bowler et~al.,}{Bowler
  et~al.}{2012}]{Bowler2012}
Bowler R. A.~A.,  et~al., 2012, \mn@doi [MNRAS]
  {10.1111/j.1365-2966.2012.21904.x}, 426, 2772

\bibitem[\protect\citeauthoryear{Bowler et~al.,}{Bowler
  et~al.}{2014}]{Bowler2014}
Bowler R. A.~A.,  et~al., 2014, \mn@doi [MNRAS] {10.1093/mnras/stu449}, 440,
  2810

\bibitem[\protect\citeauthoryear{Bowler, Dunlop, McLure  \& McLeod}{Bowler
  et~al.}{2017}]{Bowler2017}
Bowler R.,  Dunlop J.,  McLure R.,   McLeod D.,  2017, \mn@doi [MNRAS]
  {10.1093/mnras/stw3296}, 466, 3612

\bibitem[\protect\citeauthoryear{Bowler, Bourne, Dunlop, McLure  \&
  McLeod}{Bowler et~al.}{2018}]{Bowler2018}
Bowler R. A.~A.,  Bourne N.,  Dunlop J.~S.,  McLure R.~J.,   McLeod D.~J.,
  2018, \mn@doi [MNRAS] {10.1093/mnras/sty2368}, 481, 1631

\bibitem[\protect\citeauthoryear{Bruzual \& Charlot}{Bruzual \&
  Charlot}{2003}]{Bruzual2003}
Bruzual G.,  Charlot S.,  2003, \mn@doi [MNRAS]
  {10.1046/j.1365-8711.2003.06897.x}, 344, 1000

\bibitem[\protect\citeauthoryear{Buat, Ciesla, Boquien, Ma{\l}ek  \&
  Burgarella}{Buat et~al.}{2019}]{Buat2019}
Buat V.,  Ciesla L.,  Boquien M.,  Ma{\l}ek K.,   Burgarella D.,  2019, \mn@doi
  [A{\&}A] {10.1051/0004-6361/201936643}, 632, A79

\bibitem[\protect\citeauthoryear{Buitrago, Trujillo, Conselice  \&
  Haussler}{Buitrago et~al.}{2013}]{Buitrago2013}
Buitrago F.,  Trujillo I.,  Conselice C.~J.,   Haussler B.,  2013, \mn@doi
  [MNRAS] {10.1093/mnras/sts124}, 428, 1460

\bibitem[\protect\citeauthoryear{Calzetti}{Calzetti}{2001}]{Calzetti2001}
Calzetti D.,  2001, \mn@doi [PASP] {10.1086/324269}, 113, 1449

\bibitem[\protect\citeauthoryear{Calzetti, Kinney  \&
  Storchi-Bergmann}{Calzetti et~al.}{1994}]{Calzetti1994}
Calzetti D.,  Kinney A.~L.,   Storchi-Bergmann T.,  1994, \mn@doi [ApJ]
  {10.1086/174346}, 429, 582

\bibitem[\protect\citeauthoryear{Calzetti, Armus, Bohlin, Kinney, Koornneef  \&
  Storchi-Bergmann}{Calzetti et~al.}{2000}]{Calzetti2000}
Calzetti D.,  Armus L.,  Bohlin R.~C.,  Kinney A.~L.,  Koornneef J.,
  Storchi-Bergmann T.,  2000, \mn@doi [ApJ] {10.1086/308692}, 533, 682

\bibitem[\protect\citeauthoryear{Capak et~al.,}{Capak et~al.}{2015}]{Capak2015}
Capak P.~L.,  et~al., 2015, \mn@doi [Nat] {10.1038/nature14500}, 522, 455

\bibitem[\protect\citeauthoryear{Carniani et~al.,}{Carniani
  et~al.}{2017}]{Carniani2017}
Carniani S.,  et~al., 2017, \mn@doi [A{\&}A] {10.1051/0004-6361/201630366},
  605, A42

\bibitem[\protect\citeauthoryear{Carniani et~al.,}{Carniani
  et~al.}{2018}]{Carniani2018}
Carniani S.,  et~al., 2018, \mn@doi [MNRAS] {10.1093/mnras/sty1088}, 1184, 1170

\bibitem[\protect\citeauthoryear{Carniani et~al.,}{Carniani
  et~al.}{2020}]{Carniani2020}
Carniani S.,  et~al., 2020, \mn@doi [MNRAS] {10.1093/mnras/staa3178}, 499, 5136

\bibitem[\protect\citeauthoryear{Casey}{Casey}{2012}]{Casey2012}
Casey C.~M.,  2012, \mn@doi [MNRAS] {10.1111/j.1365-2966.2012.21455.x}, 425,
  3094

\bibitem[\protect\citeauthoryear{Casey et~al.,}{Casey
  et~al.}{2014}]{Casey2014a}
Casey C.~M.,  et~al., 2014, \mn@doi [ApJ] {10.1088/0004-637X/796/2/95}, 796, 95

\bibitem[\protect\citeauthoryear{Chabrier}{Chabrier}{2003}]{Chabrier2003}
Chabrier G.,  2003, PASP, 115, 763

\bibitem[\protect\citeauthoryear{Charlot \& Fall}{Charlot \&
  Fall}{2000}]{Charlot2000}
Charlot S.,  Fall S.~M.,  2000, \mn@doi [ApJ] {10.1086/309250}, 539, 718

\bibitem[\protect\citeauthoryear{Chen et~al.,}{Chen et~al.}{2015}]{Chen2015}
Chen C.~C.,  et~al., 2015, \mn@doi [AJ] {10.1088/0004-637X/799/2/194}, 799, 28

\bibitem[\protect\citeauthoryear{Chen, Hodge, Smail, Swinbank, Walter  \&
  Simpson}{Chen et~al.}{2017}]{Chen2017}
Chen C.-c.,  Hodge J.~A.,  Smail I.,  Swinbank A.~M.,  Walter F.,   Simpson
  J.~M.,  2017, \mn@doi [ApJ] {10.3847/1538-4357/aa863a}, 846, 108

\bibitem[\protect\citeauthoryear{Chen et~al.,}{Chen et~al.}{2020}]{Chen2020}
Chen C.~C.,  et~al., 2020, A{\&}A, 635, 119

\bibitem[\protect\citeauthoryear{Cheng et~al.,}{Cheng et~al.}{2020}]{Cheng2020}
Cheng C.,  et~al., 2020, \mn@doi [MNRAS] {10.1093/mnras/staa3036}, 499, 5241

\bibitem[\protect\citeauthoryear{Clark et~al.,}{Clark et~al.}{2015}]{Clark2015}
Clark C.~J.,  et~al., 2015, \mn@doi [MNRAS] {10.1093/mnras/stv1276}, 452, 397

\bibitem[\protect\citeauthoryear{Cochrane et~al.,}{Cochrane
  et~al.}{2019}]{Cochrane2019}
Cochrane R.~K.,  et~al., 2019, \mn@doi [MNRAS] {10.1093/mnras/stz1736}, 488,
  1779

\bibitem[\protect\citeauthoryear{Cochrane et~al.,}{Cochrane
  et~al.}{2021}]{Cochrane2021}
Cochrane R.~K.,  et~al., 2021, \mn@doi [MNRAS] {10.1093/MNRAS/STAB467}, 503,
  2622

\bibitem[\protect\citeauthoryear{Cullen, McLure, Khochfar, Dunlop  \&
  Vecchia}{Cullen et~al.}{2017}]{Cullen2017}
Cullen F.,  McLure R.~J.,  Khochfar S.,  Dunlop J.~S.,   Vecchia C.~D.,  2017,
  MNRAS, 470, 3006

\bibitem[\protect\citeauthoryear{Curtis-Lake et~al.,}{Curtis-Lake
  et~al.}{2016}]{Curtis-Lake2016}
Curtis-Lake E.,  et~al., 2016, \mn@doi [MNRAS] {10.1093/mnras/stv3017}, 457,
  440

\bibitem[\protect\citeauthoryear{Draine}{Draine}{2003}]{Draine2003}
Draine B.~T.,  2003, \mn@doi [ARA{\&}A]
  {10.1146/ANNUREV.ASTRO.41.011802.094840}, 41, 241

\bibitem[\protect\citeauthoryear{Duncan et~al.,}{Duncan
  et~al.}{2014}]{Duncan2014}
Duncan K.,  et~al., 2014, \mn@doi [MNRAS] {10.1093/mnras/stu1622}, 444, 2960

\bibitem[\protect\citeauthoryear{Dunlop et~al.,}{Dunlop
  et~al.}{2013}]{Dunlop2013}
Dunlop J.~S.,  et~al., 2013, \mn@doi [MNRAS] {10.1093/mnras/stt702}, 432, 3520

\bibitem[\protect\citeauthoryear{Dunlop et~al.,}{Dunlop
  et~al.}{2017}]{Dunlop2017}
Dunlop J.~S.,  et~al., 2017, \mn@doi [MNRAS] {10.1093/mnras/stw3088}, 466, 861

\bibitem[\protect\citeauthoryear{Dunne et~al.,}{Dunne et~al.}{2018}]{Dunne2018}
Dunne L.,  et~al., 2018, \mn@doi [MNRAS] {10.1093/MNRAS/STY1465}, 479, 1221

\bibitem[\protect\citeauthoryear{Faisst et~al.,}{Faisst
  et~al.}{2017}]{Faisst2017}
Faisst A.~L.,  et~al., 2017, \mn@doi [ApJ] {10.3847/1538-4357/aa886c}, 847, 28

\bibitem[\protect\citeauthoryear{Faisst et~al.,}{Faisst
  et~al.}{2020}]{Faisst2020}
Faisst A.~L.,  et~al., 2020, \mn@doi [ApJS] {10.3847/1538-4365/ab7ccd}, 247

\bibitem[\protect\citeauthoryear{{F{\"{o}}rster Schreiber}
  et~al.,}{{F{\"{o}}rster Schreiber} et~al.}{2011}]{ForsterSchreiber2011}
{F{\"{o}}rster Schreiber} N.~M.,  et~al., 2011, \mn@doi [AJ]
  {10.1088/0004-637X/739/1/45}, 739, 45

\bibitem[\protect\citeauthoryear{Fudamoto et~al.,}{Fudamoto
  et~al.}{2020a}]{Fudamoto2019}
Fudamoto Y.,  et~al., 2020a, \mn@doi [MNRAS] {10.1093/mnras/stz3248}, 491, 4724

\bibitem[\protect\citeauthoryear{Fudamoto et~al.,}{Fudamoto
  et~al.}{2020b}]{Fudamoto2020}
Fudamoto Y.,  et~al., 2020b, A{\&}A, 643

\bibitem[\protect\citeauthoryear{Fudamoto et~al.,}{Fudamoto
  et~al.}{2021}]{Fudamoto2021}
Fudamoto Y.,  et~al., 2021, \mn@doi [Nat] {10.1038/s41586-021-03846-z}, 597,
  489

\bibitem[\protect\citeauthoryear{Fujimoto, Ouchi, Shibuya  \& Nagai}{Fujimoto
  et~al.}{2017}]{Fujimoto2017}
Fujimoto S.,  Ouchi M.,  Shibuya T.,   Nagai H.,  2017, \mn@doi [ApJ]
  {10.3847/1538-4357/aa93e6}, 850, 83

\bibitem[\protect\citeauthoryear{{Gil de Paz} et~al.,}{{Gil de Paz}
  et~al.}{2007}]{GildePaz2007}
{Gil de Paz} A.,  et~al., 2007, \mn@doi [ApJSS] {10.1086/516636}, 173, 185

\bibitem[\protect\citeauthoryear{Gonz{\'{a}}lez-L{\'{o}}pez
  et~al.,}{Gonz{\'{a}}lez-L{\'{o}}pez et~al.}{2020}]{GonzalezLopez2020}
Gonz{\'{a}}lez-L{\'{o}}pez J.,  et~al., 2020, \mn@doi [ApJ]
  {10.3847/1538-4357/AB765B}, 897, 91

\bibitem[\protect\citeauthoryear{Grazian et~al.,}{Grazian
  et~al.}{2012}]{Grazian2012}
Grazian A.,  et~al., 2012, \mn@doi [A{\&}A] {10.1051/0004-6361/201219669}, 547,
  A51

\bibitem[\protect\citeauthoryear{Grogin et~al.,}{Grogin
  et~al.}{2011}]{Grogin2011}
Grogin N.~A.,  et~al., 2011, \mn@doi [ApJS] {10.1088/0067-0049/197/2/35}, 197,
  35

\bibitem[\protect\citeauthoryear{Guo et~al.,}{Guo et~al.}{2015}]{Guo2015}
Guo Y.,  et~al., 2015, \mn@doi [ApJ] {10.1088/0004-637X/800/1/39}, 800, 39

\bibitem[\protect\citeauthoryear{Hashimoto et~al.,}{Hashimoto
  et~al.}{2018}]{Hashimoto2018}
Hashimoto T.,  et~al., 2018, Nat, 557, 392

\bibitem[\protect\citeauthoryear{Hashimoto et~al.,}{Hashimoto
  et~al.}{2019}]{Hashimoto2019}
Hashimoto T.,  et~al., 2019, \mn@doi [PASJ] {10.1093/pasj/psz049}, 71, 1

\bibitem[\protect\citeauthoryear{Hodge \& da Cunha}{Hodge \&
  da~Cunha}{2020}]{Hodge2020}
Hodge J.~A.,  da Cunha E.,  2020, RSOS

\bibitem[\protect\citeauthoryear{Hodge et~al.,}{Hodge et~al.}{2016}]{Hodge2016}
Hodge J.~A.,  et~al., 2016, ApJ, 833, 103

\bibitem[\protect\citeauthoryear{Hodge et~al.,}{Hodge et~al.}{2019}]{Hodge2019}
Hodge J.~A.,  et~al., 2019, \mn@doi [ApJ] {10.3847/1538-4357/ab1846}, 876, 130

\bibitem[\protect\citeauthoryear{Kennicutt \& Evans}{Kennicutt \&
  Evans}{2012}]{Kennicutt2012}
Kennicutt R.~C.,  Evans N.~J.,  2012, \mn@doi [ARA{\&}A]
  {10.1146/annurev-astro-081811-125610}, 50, 531

\bibitem[\protect\citeauthoryear{Kennicutt et~al.,}{Kennicutt
  et~al.}{2003}]{Kennicutt2003}
Kennicutt R.,  et~al., 2003, \mn@doi [PASJ] {10.1086/376941}, 115, 928

\bibitem[\protect\citeauthoryear{Knudsen, Watson, Frayer, Christensen,
  Gallazzi, Michalowski, Richard  \& Zavala}{Knudsen
  et~al.}{2017}]{Knudsen2016}
Knudsen K.~K.,  Watson D.,  Frayer D.,  Christensen L.,  Gallazzi A.,
  Michalowski M.~J.,  Richard J.,   Zavala J.,  2017, MNRAS, 466, 138

\bibitem[\protect\citeauthoryear{Koekemoer et~al.,}{Koekemoer
  et~al.}{2011}]{Koekemoer2011}
Koekemoer A.~M.,  et~al., 2011, \mn@doi [ApJS] {10.1088/0067-0049/197/2/36},
  197, 36

\bibitem[\protect\citeauthoryear{Koprowski et~al.,}{Koprowski
  et~al.}{2016}]{Koprowski2016}
Koprowski M.~P.,  et~al., 2016, MNRAS, 828, L21

\bibitem[\protect\citeauthoryear{Koprowski et~al.,}{Koprowski
  et~al.}{2018}]{Koprowski2018}
Koprowski M.~P.,  et~al., 2018, \mn@doi [MNRAS] {10.1093/mnras/sty1527}, 479,
  4355

\bibitem[\protect\citeauthoryear{Laporte et~al.,}{Laporte
  et~al.}{2017}]{Laporte2017}
Laporte N.,  et~al., 2017, \mn@doi [ApJ] {10.3847/2041-8213/aa62aa}, 837, L21

\bibitem[\protect\citeauthoryear{Laporte et~al.,}{Laporte
  et~al.}{2021}]{Laporte2021}
Laporte N.,  et~al., 2021, \mn@doi [MNRAS] {10.1093/MNRAS/STAB191}, 505, 4838

\bibitem[\protect\citeauthoryear{Lawrence et~al.,}{Lawrence
  et~al.}{2007}]{Lawrence2007}
Lawrence A.,  et~al., 2007, \mn@doi [MNRAS] {10.1111/j.1365-2966.2007.12040.x},
  379, 1599

\bibitem[\protect\citeauthoryear{{Le F{\`{e}}vre} et~al.,}{{Le F{\`{e}}vre}
  et~al.}{2020}]{LeFevre2020}
{Le F{\`{e}}vre} O.,  et~al., 2020, \mn@doi [A{\&}A]
  {10.1051/0004-6361/201936965}, 643, A1

\bibitem[\protect\citeauthoryear{Liang, Feldmann, Hayward, Narayanan,
  {\c{C}}atmabacak, Kere{\v{s}}, Faucher-Gigu{\`{e}}re  \& Hopkins}{Liang
  et~al.}{2021}]{Liang2021}
Liang L.,  Feldmann R.,  Hayward C.~C.,  Narayanan D.,  {\c{C}}atmabacak O.,
  Kere{\v{s}} D.,  Faucher-Gigu{\`{e}}re C.-A.,   Hopkins P.~F.,  2021, \mn@doi
  [MNRAS] {10.1093/mnras/stab096}, 502, 3210

\bibitem[\protect\citeauthoryear{Madau \& Dickinson}{Madau \&
  Dickinson}{2014}]{Madau2014}
Madau P.,  Dickinson M.,  2014, \mn@doi [ARA{\&}A]
  {10.1146/annurev-astro-081811-125615}, 52, 415

\bibitem[\protect\citeauthoryear{Madau, Pozzetti  \& Dickinson}{Madau
  et~al.}{1998}]{Madau1998}
Madau P.,  Pozzetti L.,   Dickinson M.,  1998, \mn@doi [ApJ] {10.1086/305523},
  498, 106

\bibitem[\protect\citeauthoryear{Maiolino et~al.,}{Maiolino
  et~al.}{2015}]{Maiolino2015}
Maiolino R.,  et~al., 2015, MNRAS, 452, 54

\bibitem[\protect\citeauthoryear{McCracken et~al.,}{McCracken
  et~al.}{2013}]{McCracken2013}
McCracken H.~J.,  et~al., 2013, The Messenger, 154, 29

\bibitem[\protect\citeauthoryear{McLure et~al.,}{McLure
  et~al.}{2018}]{McLure2018}
McLure R.~J.,  et~al., 2018, \mn@doi [MNRAS] {10.1093/mnras/sty1213}, 479, 25

\bibitem[\protect\citeauthoryear{McMullin et~al.,}{McMullin
  et~al.}{2007}]{McMullin2007}
McMullin J.~P.,  et~al., 2007, ASP Conf. Ser., Astronomic, 376

\bibitem[\protect\citeauthoryear{Merlin et~al.,}{Merlin
  et~al.}{2015}]{Merlin2015}
Merlin E.,  et~al., 2015, \mn@doi [A{\&}A] {10.1051/0004-6361/201526471}, 582,
  A15

\bibitem[\protect\citeauthoryear{Meurer, Heckman  \& Calzetti}{Meurer
  et~al.}{1999}]{Meurer1999}
Meurer G.~R.,  Heckman T.~M.,   Calzetti D.,  1999, \mn@doi [ApJ]
  {10.1086/307523}, 521, 64

\bibitem[\protect\citeauthoryear{Micha{\l}owski et~al.,}{Micha{\l}owski
  et~al.}{2017}]{Michalowski2017}
Micha{\l}owski M.~J.,  et~al., 2017, \mn@doi [MNRAS] {10.1093/mnras/stx861},
  469, 492

\bibitem[\protect\citeauthoryear{Mohan, Rafferty, Mohan  \& Rafferty}{Mohan
  et~al.}{2015}]{Mohan2015}
Mohan N.,  Rafferty D.,  Mohan N.,   Rafferty D.,  2015, ASCL, p. ascl:1502.007

\bibitem[\protect\citeauthoryear{Oke}{Oke}{1974}]{Oke1974}
Oke J.~B.,  1974, \mn@doi [ApJS] {10.1086/190287}, 27, 21

\bibitem[\protect\citeauthoryear{Oke \& Gunn}{Oke \& Gunn}{1983}]{Oke1983}
Oke J.~B.,  Gunn J.~E.,  1983, \mn@doi [ApJ] {10.1086/160817}, 266, 713

\bibitem[\protect\citeauthoryear{Ouchi et~al.,}{Ouchi et~al.}{2013}]{Ouchi2013}
Ouchi M.,  et~al., 2013, \mn@doi [ApJ] {10.1088/0004-637X/778/2/102}, 778, 102

\bibitem[\protect\citeauthoryear{Pelliccia et~al.,}{Pelliccia
  et~al.}{2021}]{Pelliccia2021}
Pelliccia D.,  et~al., 2021, \mn@doi [ApJL] {10.3847/2041-8213/abdf56}, 908,
  L30

\bibitem[\protect\citeauthoryear{Popping, Somerville  \& Galametz}{Popping
  et~al.}{2017}]{Popping2017}
Popping G.,  Somerville R.~S.,   Galametz M.,  2017, \mn@doi [MNRAS]
  {10.1093/mnras/stx1545}, 471, 3152

\bibitem[\protect\citeauthoryear{Popping et~al.,}{Popping
  et~al.}{2021}]{Popping2021}
Popping G.,  et~al., 2021, preprint (arXiv:2021.12218)

\bibitem[\protect\citeauthoryear{Rivera et~al.,}{Rivera
  et~al.}{2018}]{Rivera2018a}
Rivera G.~C.,  et~al., 2018, \mn@doi [ApJ] {10.3847/1538-4357/aacffa}, 863, 56

\bibitem[\protect\citeauthoryear{Rogers, McLure  \& Dunlop}{Rogers
  et~al.}{2013}]{Rogers2013}
Rogers A.~B.,  McLure R.~J.,   Dunlop J.~S.,  2013, \mn@doi [MNRAS]
  {10.1093/mnras/sts515}, 429, 2456

\bibitem[\protect\citeauthoryear{Rujopakarn et~al.,}{Rujopakarn
  et~al.}{2019}]{Rujopakarn2019}
Rujopakarn W.,  et~al., 2019, \mn@doi [ApJ] {10.3847/1538-4357/ab3791}, 882,
  107

\bibitem[\protect\citeauthoryear{Schouws et~al.,}{Schouws
  et~al.}{2021}]{Schouws2021}
Schouws S.,  et~al., 2021, preprint (arXiv:2105.12133)

\bibitem[\protect\citeauthoryear{Scoville et~al.,}{Scoville
  et~al.}{2007}]{Scoville2007}
Scoville N.,  et~al., 2007, \mn@doi [ApJS] {10.1086/516585}, 172, 1

\bibitem[\protect\citeauthoryear{Smit et~al.,}{Smit et~al.}{2018}]{Smit2018a}
Smit R.,  et~al., 2018, \mn@doi [Nat] {10.1038/NATURE24631}, 553, 178

\bibitem[\protect\citeauthoryear{Sommovigo, Ferrara, Carniani, Zanella,
  Pallottini, Gallerani  \& Vallini}{Sommovigo et~al.}{2021}]{Sommovigo2021}
Sommovigo L.,  Ferrara A.,  Carniani S.,  Zanella A.,  Pallottini A.,
  Gallerani S.,   Vallini L.,  2021, \mn@doi [MNRAS] {10.1093/mnras/stab720},
  503, 4878

\bibitem[\protect\citeauthoryear{Stark, Schenker, Ellis, Robertson, McLure  \&
  Dunlop}{Stark et~al.}{2013}]{Stark2013}
Stark D.~P.,  Schenker M.~A.,  Ellis R.,  Robertson B.,  McLure R.,   Dunlop
  J.,  2013, \mn@doi [ApJ] {10.1088/0004-637X/763/2/129}, 763, 129

\bibitem[\protect\citeauthoryear{Steidel, Bogosavljevi{\'{c}}, Shapley,
  Kollmeier, Reddy, Erb  \& Pettini}{Steidel et~al.}{2011}]{Steidel2011}
Steidel C.~C.,  Bogosavljevi{\'{c}} M.,  Shapley A.~E.,  Kollmeier J.~A.,
  Reddy N.~A.,  Erb D.~K.,   Pettini M.,  2011, \mn@doi [ApJ]
  {10.1088/0004-637X/736/2/160}, 736, 160

\bibitem[\protect\citeauthoryear{Sugahara et~al.,}{Sugahara
  et~al.}{2021}]{Sugahara2021}
Sugahara Y.,  et~al., 2021, preprint (arXiv:2104.02201)

\bibitem[\protect\citeauthoryear{Tadaki et~al.,}{Tadaki
  et~al.}{2020}]{Tadaki2020}
Tadaki K.-i.,  et~al., 2020, \mn@doi [ApJ] {10.3847/1538-4357/abaf4a}, 901, 74

\bibitem[\protect\citeauthoryear{Takeuchi, Yuan, Ikeyama, Murata  \&
  Inoue}{Takeuchi et~al.}{2012}]{Takeuchi2012}
Takeuchi T.~T.,  Yuan F.-T.,  Ikeyama A.,  Murata K.~L.,   Inoue A.~K.,  2012,
  \mn@doi [ApJ] {10.1088/0004-637X/755/2/144}, 755, 144

\bibitem[\protect\citeauthoryear{Tamura et~al.,}{Tamura
  et~al.}{2019}]{Tamura2019}
Tamura Y.,  et~al., 2019, \mn@doi [ApJ] {10.3847/1538-4357/ab0374}, 874, 27

\bibitem[\protect\citeauthoryear{Targett et~al.,}{Targett
  et~al.}{2013}]{Targett2013}
Targett T.~A.,  et~al., 2013, \mn@doi [MNRAS] {10.1093/mnras/stt482}, 432, 2012

\bibitem[\protect\citeauthoryear{Watson, Christensen, Knudsen, Richard,
  Gallazzi  \& Micha{\l}owski}{Watson et~al.}{2015}]{Watson2015}
Watson D.,  Christensen L.,  Knudsen K.~K.,  Richard J.,  Gallazzi A.,
  Micha{\l}owski M.~J.,  2015, \mn@doi [Nat] {10.1038/nature14164}, 519, 327

\bibitem[\protect\citeauthoryear{Wuyts et~al.,}{Wuyts et~al.}{2012}]{Wuyts2012}
Wuyts S.,  et~al., 2012, \mn@doi [ApJ] {10.1088/0004-637X/753/2/114}, 753, 114

\makeatother
\end{thebibliography}



\bsp	
\label{lastpage}
\end{document}